\documentclass[12pt,a4paper]{article}
\usepackage{graphicx}
\usepackage{color}
\usepackage{cite}
\usepackage{amsmath,amssymb}

\newcommand{\veps}{\varepsilon}
\newcommand{\vphi}{\varphi}

\newcommand{\bit}{\begin{itemize}}
\newcommand{\eit}{\end{itemize}}
\newcommand{\ben}{\begin{enumerate}}
\newcommand{\een}{\end{enumerate}}
\newcommand{\beq}{\begin{equation}}
\newcommand{\eeq}{\end{equation}}
\newcommand{\beqa}{\begin{eqnarray}}
\newcommand{\eeqa}{\end{eqnarray}}
\newcommand{\beqn}{\begin{eqnarray*}}
\newcommand{\eeqn}{\end{eqnarray*}}

\newcommand{\corrL}{}
\newcommand{\corrS}{}

\begin{document}

\title{Periodic traveling waves in a taut cable on a bilinear elastic substrate}
\date{}
\author{{\bf Lucio Demeio} \\
Dipartimento di Scienze Matematiche \\
Universit\`a Politecnica delle Marche \\
Via Brecce Bianche 1 \\
I-60131 Ancona, Italy \\
(corresponding author) \\
e-mail: demeio@mta01.univpm.it \\

\vspace*{0.2 cm}
and \\

\vspace*{0.2 cm}
{\bf Stefano Lenci}\\
Dipartimento di Architettura, Costruzioni e Strutture  \\
Universit\`a Politecnica delle Marche \\
Via Brecce Bianche 1 \\
I-60131 Ancona, Italy \\
e-mail: lenci@univpm.it }

\maketitle

\begin{abstract}
  The wave propagation problem on a taut cable resting on a bilinear substrate is investigated, without and with a distribute transversal load. {\corrL The  piecewise nature of the problem offers a sufficiently simple kind of nonlinearity as to permit a closed form solution} both for the wave phase velocity and the wave form. We show that the solution depends only on the ratio between the two soil stiffnesses, and that no waves propagate if one side of the substrate is rigid. Some numerical simulations, based on a finite difference method, are performed to confirm the analytical findings. The stability of the proposed waves is discussed analytically and numerically.
\end{abstract}

\section{Introduction}
\label{introduction}

Within the realm of nonlinear problems in science and engineering, an important role is played by the piecewise linear systems \cite{Maez,AnViKh,Fars,DBBuChKo}. In fact, from the one side, they have all the characteristics, and complexity, of nonlinear models, with a cornucopia of cases and phenomena to be avoided or exploited, many more than in the linear case. From the other side, they often possess closed form solutions, which are useful for the full understanding and for the possibility of performing detailed parametric analyses, seldom feasible by numerical means \cite{ShHo,ThGh,FrPoRoTo}. Also, sometimes the piecewise linear system has a richer dynamics than the smooth nonlinear counterpart \cite{WoKo,CaFrPoTo}. \\

Among the many examples of piecewise linear systems in engineering applications, we mention the  inverted pendulum with lateral barriers \cite{LeRe1,LeDePe}, oscillations of rigid blocks \cite{LeRe2}, vibro-impact systems \cite{PuPe,StDAAn}, cracked beams \cite{ShPi,Mora,CeVe} (where the crack is treated as a bilinear lumped spring), capsule systems \cite{YaLiLi}, drillstring problems \cite{DiSaWiPa} and neuronal networks with gap junctions \cite{Co}. 
Referring to continuous systems, relevant examples are those of beams and cables resting on unilateral foundations \cite{DeLe,TsWe,CeMa,BhBa}. \\

In this paper we consider the wave propagation in a taut cable, or string, resting on a piecewise linear foundation, which is a paradigmatic example of a continuous bilinear system {\corrL and contains a simple form of nonlinearity. In mathematical language the wave equation with the addition of a restoring term is known as the Klein-Gordon equation \cite{Scot}, and arises, among others, in quantum mechanics. When the restoring term is linear in the displacement we obtain the linear Klein-Gordon equation, while other shapes of the restoring term, including a piecewise linear expression, lead to a nonlinear Klein-Gordon equation. As is well known, the same equation holds for axial and torsional wave propagating in beams and in other physical systems}. \\

Although wave propagation in taut cables is a well-known and well investigated problem \cite{Ko,Gr}, even in the presence of an elastic foundation, to the best of our knowledge the problem with a piecewise linear distributed support has not been previously addressed, and we hope to fill this gap. 
In fact, while the dynamics of beams resting on bilinear foundations has been investigated by various authors \cite{FaSh,JoKo,Adin,LiWaPa,CaPi,Mazi1,Froio,FrRi,ZhLiWe,Lenc}, to the best of our knowledge a similar analysis has not been done for the string, in spite of the simpler equation involved. 
For example, the recent extended review by Younesian et al. \cite{YoHoAsEs} does not mention this type of foundation, although it considers other nonlinear substrate models and although it affirms that ``a foundation model may present different behavior under tension and pressure loads". As a matter of fact, this work extends to strings the same problem studied in \cite{Lenc} for the beam. \\

A particular case is obtained when one of the two stiffnesses is zero, which corresponds to a unilateral substrate. This situation has been considered in \cite{Metr} for the case of  two moving lumped forces, although the problem of free wave propagation is not addressed directly.
This problem finds application, for example, in railway overhead power lines. Actually, this engineering example constitutes one of the practical problems which triggered this study. The case of a cable in between two different media (for example in underground power lines) or that of filaments resting on two different tissues (in biomechanics) and other interface problems are further examples. Referring to the engineering case of axial wave propagation in beams, it finds applications in foundation piles, since in general the surrounding soil does not have a symmetrical behaviour. \\

The findings of this paper can be potentially used in the field of architectured materials, or metamaterials, where the material is properly designed and tailored to obtain specific wave propagation properties, related to the loss of symmetrical behaviour \cite{Oliv}. Another element of novelty of the present work is that the stability problem is addressed. Again, the stability has been investigated for beams in \cite{Lenc} with a technique similar to the one herein employed, and in \cite{NaPlMc}, although in this latter work the bilinear substrated has been approximated by a smoother function, but not for strings. This problem is challenging, and will be addressed analytically only for a subset of possible perturbations around the considered propagating waves, and numerically for more generic perturbations. \\

\begin{figure}
  \centering
a)  \includegraphics[width=8cm]{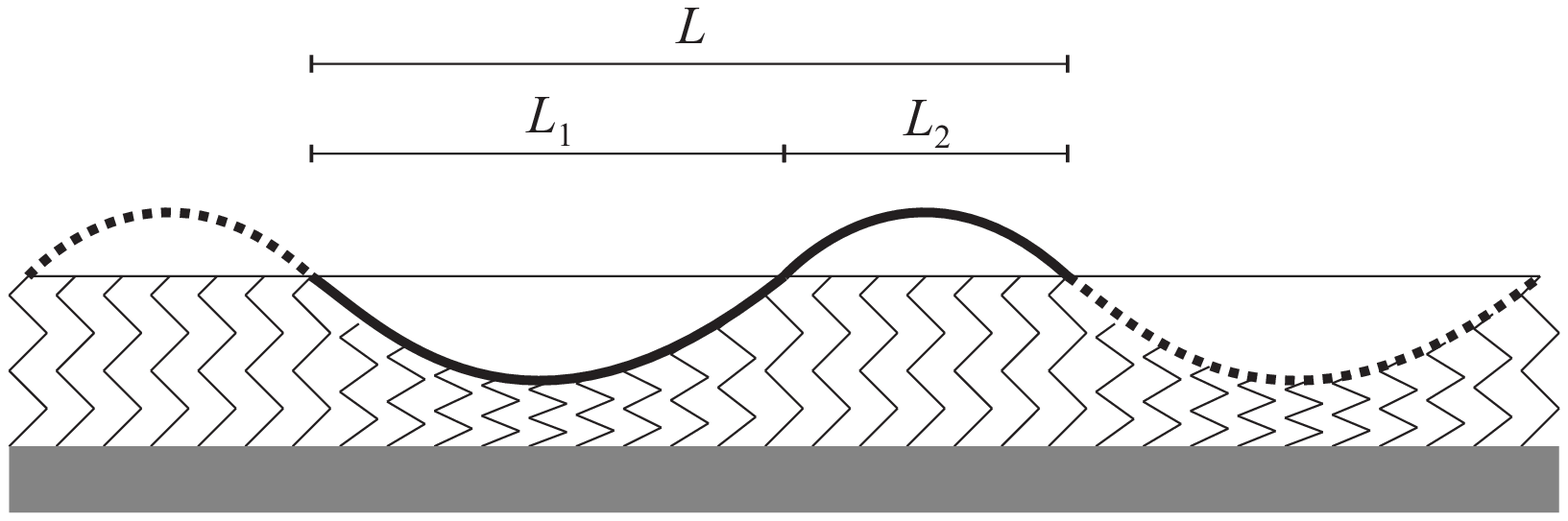} \\
\vspace{0.5cm}
b)  \includegraphics[width=11cm]{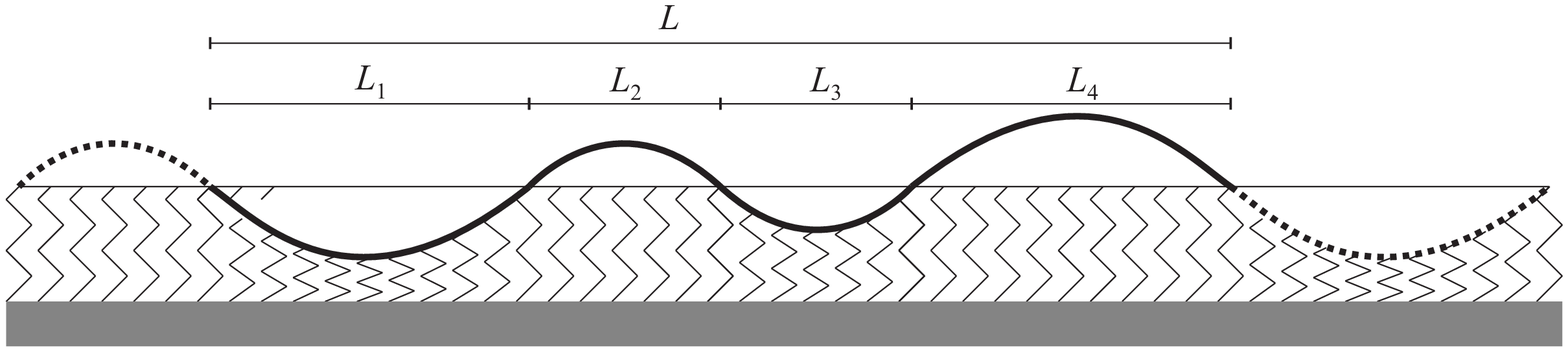}
  \caption{a) Single wave and b) multiple wave travelling periodic solutions.}
  \label{fig01}
\end{figure}

We consider traveling waves of wavelength $L$ and whose shape changes sign at least once within the wavelength. If there is only one change of sign within the wave length $L$ we name it ``single wave" (Fig. \ref{fig01}a); this is the solution of main interest in this work. If the change of sign occurs in more than one point within the wavelength we name it ``multiple wave" (Fig. \ref{fig01}b). \\

By repeating $n$ times a single wave solution we (trivially) obtain a multiple wave solution of wave length $nL$, which we name ``repetitive multiple wave". Although it represents the same physical wave, it is a different solutions of the mathematical problem which we will introduce in our analytical computations, because we will use the wavelength $L$ to define a nondimensional space variable. Thus, repetitive multiple waves must be considered in  computations, even if they do not represent new physical solutions and the only interesting case is with $n=1$. \\

The paper is organized as follows: in Section \ref{cable} we introduce the problem,
the equations and
the periodic-wave analytical solutions; in Section \ref{sec:numerical} we present some numerical simulations and Section \ref{conclusions} contains our conclusions.


\section{Governing equations and analytical solutions}
\label{cable}
The governing equation for a taut cable on a general nonlinear substrate and without external excitation is given by the nonlinear Klein-Gordon equation {\corrL \cite{gravel,shatah}}
\beq
\frac{\partial^2 w}{\partial t^2}- v^2 \, \frac{\partial^2 w}{\partial x^2}+ \gamma(w) \, w =0, \label{waveeq}
\eeq
where $w(x,t)$ is the cable profile 
with respect
to the substrate, $x \in \mathbb{R}$ is the space variable, $t \ge 0$ the time, $v=\sqrt{T/\rho}$ is the wave phase speed in absence of the substrate ($\rho$ is the mass per unit length and $T$ the tension of the cable), and $\gamma(w)$ is the stiffness of the substrate. For different mechanical problems, governed by the same equation, the wave speed  $v$ has different expressions.\\

We shall investigate the existence and the functional form of traveling wave solutions of equation (\ref{waveeq}) for the case of a bilinear substrate, i.e. for a piecewise constant stiffness function $\gamma(w)$. For the purpose of the numerical simulations, we shall also consider the associated initial/boundary value problem 
\beqa
&& \frac{\partial^2 w}{\partial t^2}  - v^2 \frac{\partial^2 w}{\partial x^2} +\gamma(w) \, w = 0, \quad x \ge 0  \label{kgnl}, \\
&& w(0,t) = \vphi (t),  \label{kgnlbc} \\
&& w(x,0) = w_0(x), \qquad \frac{\partial w}{\partial t} (x,0) = \psi(x),  \label{kgnlic}
\eeqa
where $\vphi (t)$, $w_0(x)$ and $\psi(x)$ are given functions. When the initial value problem given by equations (\ref{kgnl}) and (\ref{kgnlic}) is posed over the whole real line, the total energy is conserved if the initial condition $(w_0(x),\psi(x))$ has  compact support. If the problem is considered on the half-space $x \in \mathbb{R}^+$ with the boundary condition imposed in (\ref{kgnlbc}) the total energy is not conserved, but it is easy to recover an expression for its time dependence. Let 
\beqa
&& T(t) = \int_0^{\infty} \frac{1}{2} \, \left( \frac{\partial w}{\partial t}\right)^2 \, dx,  \\
&& V(t) = \int_0^{\infty} \left\{ \frac{1}{2} \, \left(v \, \frac{\partial w}{\partial x}\right)^2 + \Gamma(w) \right\} \, dx, \\
&& E(t) = T(t)+V(t),
\eeqa
where $\Gamma(w)$ is a primitive of $\gamma(w) \, w$, be the kinetic energy, the potential energy and the total energy for the half-space problem.
We then have:
\beqa
&& \frac{d E}{d t} = \frac{d }{d t} \, \int_0^{\infty} \left\{ \frac{1}{2} \, \left( \frac{\partial w}{\partial t}\right)^2 + \frac{1}{2} \, \left(v \, \frac{\partial w}{\partial x}\right)^2  +\Gamma(w) \right\}  \, dx = \nonumber \\
&& \quad = \int_0^{\infty} \left\{ \dot{w} \, \ddot{w} + v^2 \, w' \, \dot{w}'  + \gamma(w) \, w \, \dot{w} \right\}  \, dx = \nonumber \\
&& \quad =  \lim_{R \to \infty} \int_0^R \left\{ \dot{w} \, \ddot{w} + v^2 \, w' \, \dot{w}' + \gamma(w) \, w \, \dot{w} \right\}  \, dx. \phantom{+++}
\eeqa
After integrating the second term by parts and assuming that the initial condition has compact support we have:
\beqa
&& \frac{d E}{d t} = \lim_{R \to \infty} \left\{ \int_0^R \left( \dot{w} \, \ddot{w} - v^2 \, w'' \, \dot{w} + \gamma(w) \, w \, \dot{w} \right)  \, dx + v^2 \, [w' \, \dot{w}]_0^R \right\} = \nonumber \\
&& \quad = \lim_{R \to \infty} \left\{ \int_0^R \dot{w} \, \left( \ddot{w} - v^2 \, w'' + \gamma(w) \, w \right)  \, dx + v^2 \, [w' \, \dot{w}]_0^R \right\} = \nonumber \\
&& \quad = - v^2 \, \frac{\partial w}{\partial x}(0,t) \, \frac{\partial w}{\partial t}(0,t) = -v^2 \, \dot{\vphi}(t) \, \frac{\partial w}{\partial x}(0,t), \label{dEdt}
\eeqa
where it has been taken into account that
$\ddot{w} - v^2 \, w'' + \gamma(w) \, w = 0$ 
because of (\ref{kgnl}) and that 
$$
\lim_{R \to \infty} \, \frac{\partial w}{\partial x}(R,t) \, \frac{\partial w}{\partial t}(R,t) =0
$$
because of the compactness of the support of the initial condition. We have used equation (\ref{dEdt}) as a benchmark in some of our numerical simulations. \\

\subsection{Periodic solutions} \label{ssec:periodic}
We look for traveling wave  periodic solutions of equation (\ref{waveeq}), that is continuously differentiable solutions of the form
\beq
w(x,t) = W(s),  \label{eq02}
\eeq
with $s=x-\hat{c} \, t$; for convenience, we  introduce the nondimensional variables
\beq
c= \frac{\hat{c}}{v}, \quad k(W)=\gamma (W) \, \frac{L^2}{v^2}, \quad \xi=\frac{s}{L}, \label{eq03}
\eeq
 where $L$ is the wavelength. By substituting equations (\ref{eq02}) and (\ref{eq03}) in (\ref{waveeq}) we obtain the following ordinary differential equation for $W(\xi)$
\beq
(c^2-1) \, W''(\xi) + k (W) \, W(\xi) = 0, \label{eqW}
\eeq
whose solution for $\xi \in [0,1]$ provides the functional form of the traveling wave within one wavelength. \\

In this work, the substrate stiffness is represented by the  piecewise constant function
\beqa
&& k(W) = k_1, \quad W \le 0 \label{stiffness1} \\
&& k(W) = k_2, \quad W > 0 \label{stiffness2}
\eeqa
for real positive constants $k_1$ and $k_2$. Accordingly, let $W_1$ and $W_2$ be the restrictions of the profile function $W(\xi)$ to the intervals characterized by the sign of $W$, namely $W(\xi) = W_1(\xi)$ if $W \le 0$  (compression interval) and $W(\xi) = W_2(\xi)$ if $W > 0$  (tension interval). Note that, by the symmetry of the equations, if $W(\xi)$ is a solution corresponding to a propagation velocity $c$ for given values of the stiffnesses $k_1$ and $k_2$, then $-W(\xi)$ is the solution with the same propagating speed $c$ and the stiffnesses reversed ($k_1 \to k_2$, $k_2 \to k_1$). This allows us to consider only the case $k_1 \leq k_2$. \\

With (\ref{stiffness1}) and (\ref{stiffness2}) equation (\ref{eqW}) 
becomes a system of two differential equations for the two functions $W_1$ and $W_2$,
\begin{align}
& (c^2-1) \frac{d^2 W_1}{d \xi^2}(\xi) + k_1 W_1(\xi)=0, \quad \mathrm{for} \quad W_1(\xi) \leq 0, \label{eq04a} \\
& (c^2-1) \frac{d^2 W_2}{d \xi^2}(\xi) + k_2 W_2(\xi)=0, \quad \mathrm{for} \quad W_2(\xi) \, {\corrS >} \, 0. \label{eq04b}
\end{align}

Equation (\ref{eqW}) or the equivalent  equations (\ref{eq04a})-(\ref{eq04b})  form  a nonlinear eigenvalue problem of the second type \cite{chiappinelli,betke,berger}. For any given value of $k_1$ and $k_2$  we expect a multiplicity of solutions, some (often only one) of which correspond to the single wave solutions under investigation, while others are expected to be multiple waves. \\

As anticipated in the introduction, we are interested in 
 single
wave solutions which change sign only once in the interval $(0,1)$; with an appropriate choice of the reference frame, this is equivalent to say that there exists 
 one
$\alpha \in (0,1)$ such that $W(0) = W(\alpha) = W(1) = 0$, $W(\xi) = W_1(\xi) < 0$ for $0 < \xi < \alpha$ and $W(\xi) = W_2(\xi) >0$ for 
$\alpha < \xi < 1$.
The dimensionless spatial extension of the two  intervals are then $\alpha$ and $1-\alpha$, their physical extensions (in the $s$ variable) being $L_1=\alpha L$ and $L_2=\left( 1 - \alpha \right) L$, respectively. The wavelength $L$ is then a free  scaling parameter, while $L_1$ and $L_2$ have to be determined as part of the solution. Thus, this is a free boundary problem.

With
\begin{align}
&a^2 =\frac{k_1}{c^2-1}, \quad b^2= \frac{k_2}{c^2-1}, \label{eq05}
\end{align}
the solution of (\ref{eq04a})-(\ref{eq04b}) can be written in the form
\begin{align}
&W_1(\xi)=c_1 \sin \left( a \xi \right) + c_2 \cos \left( a \xi \right), \label{eq06a} \\
&W_2(\xi)=c_3 \sin \left( b (\xi-\alpha) \right) + c_4 \cos \left( b (\xi-\alpha) \right), \label{eq06b}
\end{align}
which, together with (\ref{eq02}), implies that only mono-harmonic waves can propagate. The solution of equations (\ref{eq04a})-(\ref{eq04b}) is determined up to an arbitrary multiplicative constant; one of the four constants $c_1$, $c_2$, $c_3$, $c_4$ will therefore remain undetermined. Note that $b=a \sqrt{k_2/k_1}$, and that $a$ and $b$ must be real otherwise the solution would be hyperbolic and it would not be possible to fulfill the boundary conditions with a non-trivial solution. This entails $c>1$, i.e. the physical phase velocity $\hat{c}$ must be greater than $v$, as expected since with  the soil the system is stiffer and the velocity of wave propagation is larger. \\

The boundary conditions on $W_1$ and $W_2$ at $\xi=0$, $\xi=\alpha$ and $\xi=1$ are given by the continuity of the function $W$ and of its slope. Continuity and periodicity of the function gives 
\begin{align}
& W_1(0)=0,  \label{eq07a} \\
& W_1(\alpha)=0,  \label{eq07b} \\
& W_2(\alpha)=0, \label{eq07c} \\
& W_2\left( 1 \right)=0, \label{eq07d}
\end{align}
which entail $c_2=c_4=0$, and
\beq
 a \, \alpha = \pi \quad \mbox{and} \quad  b(1-\alpha) = \pi. \label{eq07e}  
\eeq
By substituting equations (\ref{eq07e}) into  (\ref{eq05}) we obtain the following relationships between the phase velocity $c$ and the parameter $\alpha$:
\beqa
&& c^2=1 + \frac{k_{1} \alpha^2}{\pi^2}, \label{eq08a} \\
&& c^2=1 + \frac{k_2 \left(1-\alpha\right)^2}{\pi^2}, \label{eq08b}
\eeqa
from which we can determine $\alpha$ and $c$ in terms of $k_1$ and $k_2$ for 
single
traveling waves:
\beqa
&& \alpha= \frac{\sqrt{k_2}}{\sqrt{k_1}+\sqrt{k_2}} = \frac{1}{1+\sqrt{k_1/k_2}} \label{alpha} \\
&& c^2 = 1+\frac{1}{\pi^2} \, \frac{k_1 \, k_2}{(\sqrt{k_1}+\sqrt{k_2})^2}. \label{phasec}
\eeqa
We see that $\alpha$ (and therefore the shape of the solution)
is a function only of the ratio $k_1 / k_2$ and not of $k_1$ and $k_2$ separately. \\

For the sake of completeness, we also report the boundary conditions in the  case of a repetitive multiple wave solution; only equation (\ref{eq07d}) would change into 
\beq
W_2\left( \frac{1}{n} \right)=0
\eeq
which implies
\beqa
&& \alpha= \frac{1}{n} \, \frac{\sqrt{k_2}}{\sqrt{k_1}+\sqrt{k_2}} = \frac{1}{n} \, \frac{1}{1+\sqrt{k_1/k_2}}, \label{alphan} \\
&& c^2 = 1+\frac{1}{(n \, \pi)^2} \, \frac{k_1 \, k_2}{(\sqrt{k_1}+\sqrt{k_2})^2}. \label{phasecn}
\eeqa
Note that in principle other multiple wave solutions, \emph{non repetitive}, are possible. Expressions (\ref{alphan}) and (\ref{phasecn}) show that the eigenvalues form a countably infinite set. \\

Returning to the 
single wave case, the boundary conditions on the continuity of the slopes are given by
\beqa
&& W'_1(0)=W_2' \left(1\right), \label{eq11} \\
&& W'_1(\alpha)=W_2'(\alpha), \label{eq12}
\eeqa
which entail $c_1=- c_3 \sqrt{k_2/k_1}$ and  equation (\ref{eq12}) is  automatically satisfied (this can be easily understood from geometrical considerations, since the slope of the sine function at $\pi$ is opposite to the slope at $0$). \\

By collecting all results, the solution of equations  (\ref{eq04a})-(\ref{eq04b}) becomes:
\begin{align}
&W_1(\xi)=- c_3 \sqrt{\frac{k_2}{k_1}} \sin \left(  \frac{\xi \pi}{\alpha}  \right), \quad 0 \leq \xi \leq \alpha, \label{eq13a} \\
&W_2(\xi)=c_3 \sin \left( \frac{(\xi-\alpha) \,  \pi}{1-\alpha} \right), \quad \alpha \leq \xi \leq 1, \label{eq13b}
\end{align}
where $\alpha$ is given by (\ref{alpha}), and $c_3$ is the undetermined amplitude of the wave which can be chosen by some normalization. The period and the frequency of the wave are given by
\beq
\tau = \frac{L}{c \, v}, \qquad \omega = \frac{2 \, \pi}{\tau}.  \label{frequency}
\eeq

\subsection{Stability}\label{ssec:stability}
The stability of periodic solutions with a bilinear stiffness term has been studied only with regard to specific situations or with drastic model assumptions which essentially change the nature of the problem in the context of the beam equation \cite{NaPlMc}. In \cite{shatah} the stability of the solutions for the Klein-Gordon equation has been addressed in a quantum-mechanical context with a regular stiffness term. Here, we provide some basic results on the issue, by using some general theoretical considerations (see here below) and by numerical means (see section \ref{sec:numerical}), without the pretense of giving an exhaustive and complete view. \\

It is easy to see that the periodic solutions of equations (\ref{eq04a})-(\ref{eq04b}) are 
\emph{orbitally stable (but not stable in the Lyapunov sense nor asymptotically stable) with respect to  perturbations of the form given by (\ref{eq02})}. First of all, we note that the phase portrait $(W(\xi),W'(\xi))$
$0 \le \xi \le 1$, of the solutions of (\ref{eq04a})-(\ref{eq04b}) is made of two half ellipses, of different vertical semi-axis and connected at $W_1=W_2=0$. The previous statement can be made more precise, still remaining within the perturbations given by (\ref{eq02}), by regarding the system (\ref{eq04a})-(\ref{eq04b}) as a map from $\xi=0$ to $\xi=1$;  our periodic solution is then a fixed point of this map at $(W(0),W'(0))=(0,-c_3 \sqrt{\frac{k_2}{k_1}} \frac{\pi}{\alpha})$. The linearization of the map near the fixed point can easily be constructed by considering a small perturbation to the initial condition at $\xi=0$. The Jacobian of the map at the periodic orbit can then be calculated and it easy to see that the two Floquet multipliers are both equal to $+1$. This proves the orbital stability 
with respect to the perturbations of the form given by (\ref{eq02}). \\

More detailed investigations on the stability of the periodic solutions of equation (\ref{eq04a})-(\ref{eq04b}) will be carried out by numerical means in section \ref{sec:numerical}, both with respect to perturbations on the boundary conditions (\ref{kgnlbc}) and on the initial conditions (\ref{kgnlic}).

\subsection{Particular cases}
When $k_1=k_2=k$ the substrate is linear and the solution is well known \cite{Ko,Gr}. We have
\begin{align}
\alpha = \frac{1}{2}, \quad c^2=1+\frac{k}{(2\pi)^2}, \label{eq16}
\end{align}
and $W(\xi)$ is a simple sine function.

\subsubsection{Unilateral substrate} \label{unilateral}
\begin{figure}
  \centering
a)  \includegraphics[width=6cm]{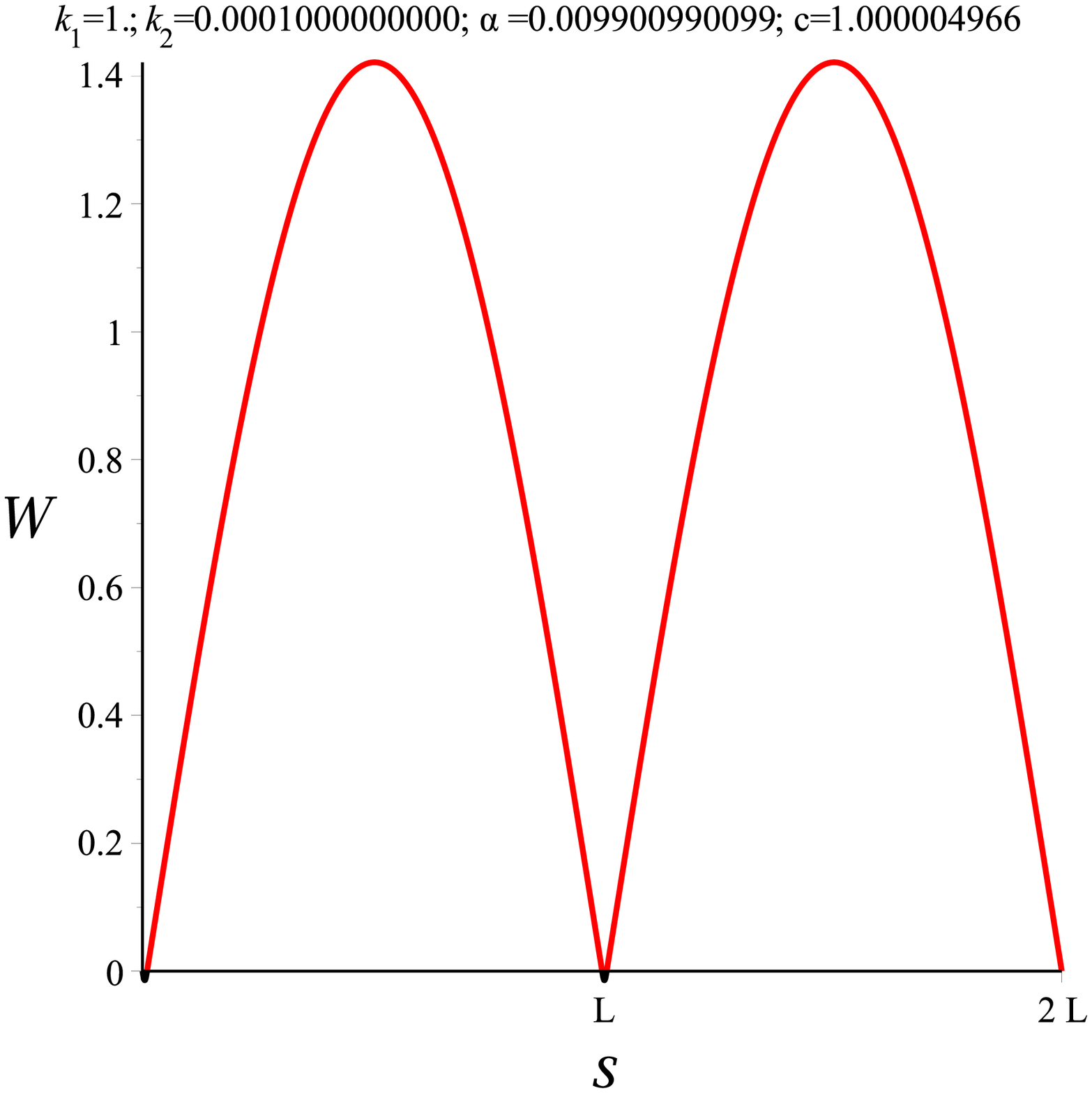}
b)  \includegraphics[width=6cm]{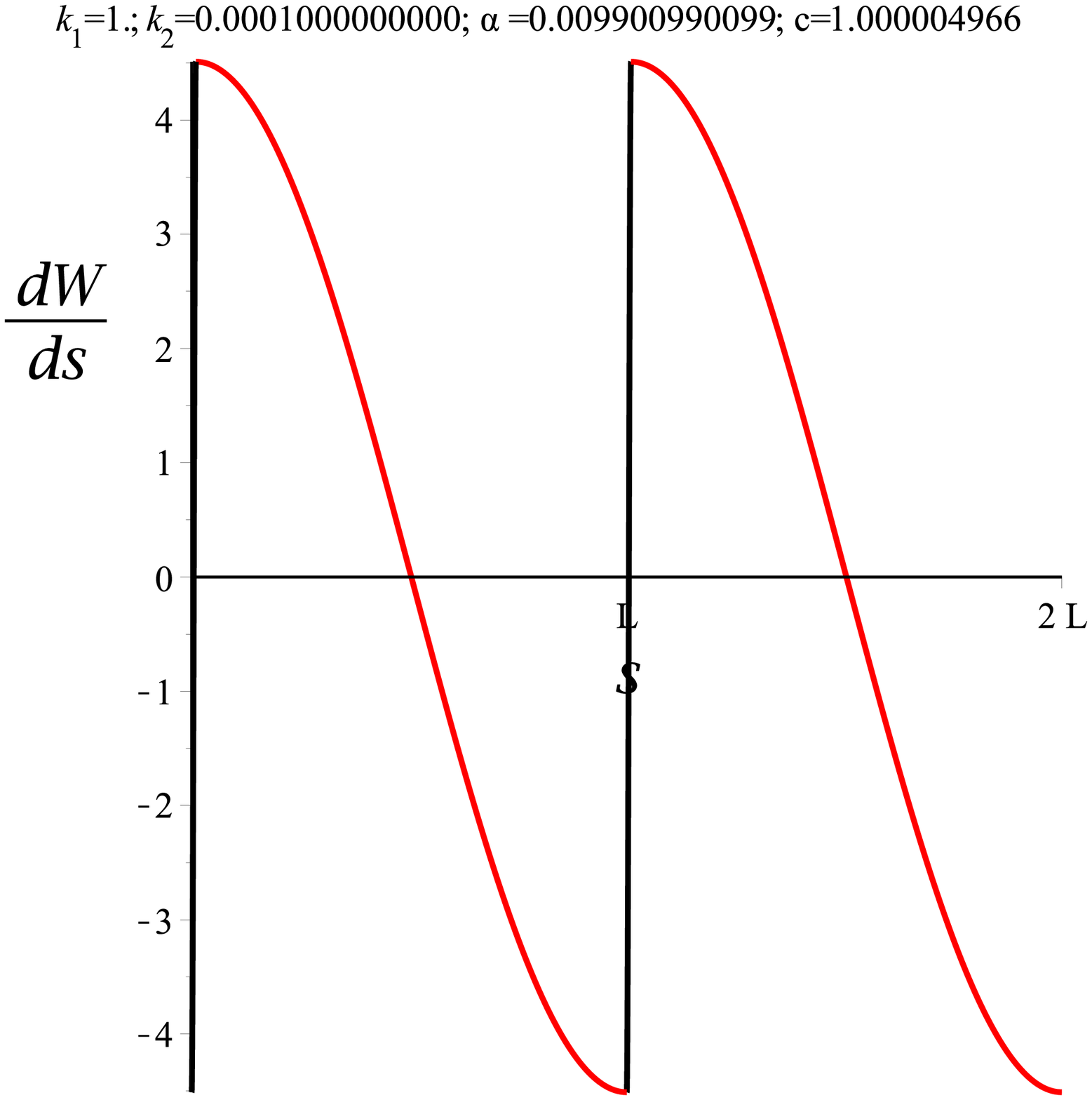}
  \caption{The solution $W(s)$ for $k_1=1$ and $k_2=10^{-4}$. $n=1$.}
  \label{fig02}
\end{figure}

When $k_2 \rightarrow 0$, the substrate becomes unilateral. In this limit we have  $\alpha \rightarrow 0$, $c \rightarrow 1$ and $W_1(\xi) \rightarrow 0$ (see (\ref{alpha}), (\ref{phasec}), (\ref{eq13a}) and (\ref{eq13b})),  which means that the wave propagates with the same speed as in the absence of the substrate and the compression region reduces to one point ($s=0$). An example of the solution over a two-period interval is reported in Fig. \ref{fig02}a, from which it is possible to see that the solution remains in the tension part. In addition, the derivative has a jump (Fig. \ref{fig02}b), since
\begin{align}
\lim_{k_2 \rightarrow 0} W_1'(\xi) \approx \lim_{k_2 \rightarrow 0} \cos \left( \frac{\pi \xi n \sqrt{k_1}}{\sqrt{k_2}} \right) \label{eq17}
\end{align}
is undefined, although bounded in a neighborhood of $k_2=0$.

The conclusion is that periodic waves  with regular profile on a perfectly unilateral substrate, crossing the region $w<0$ (where $k_1>0$)  do not exist. Of course, in this case waves of arbitrary shape (because of the D'Alembert solution) can propagate remaining always in the detached part $w>0$. \\

This result can be  interpreted also from a mechanical point of view. In fact, in the region $w>0$ ($k_2=0$) the wave propagates with velocity $c=1$  (see (\ref{eq08b})), while in the region $w<0$ ($k_1>0$) it would propagate with velocity $c>1$  (see (\ref{eq08a})), and thus it is not possible to match them.

\subsubsection{Unilaterally rigid substrate} \label{sec:unilaterallyrigid}
\begin{figure}
  \centering
a)  \includegraphics[width=6cm]{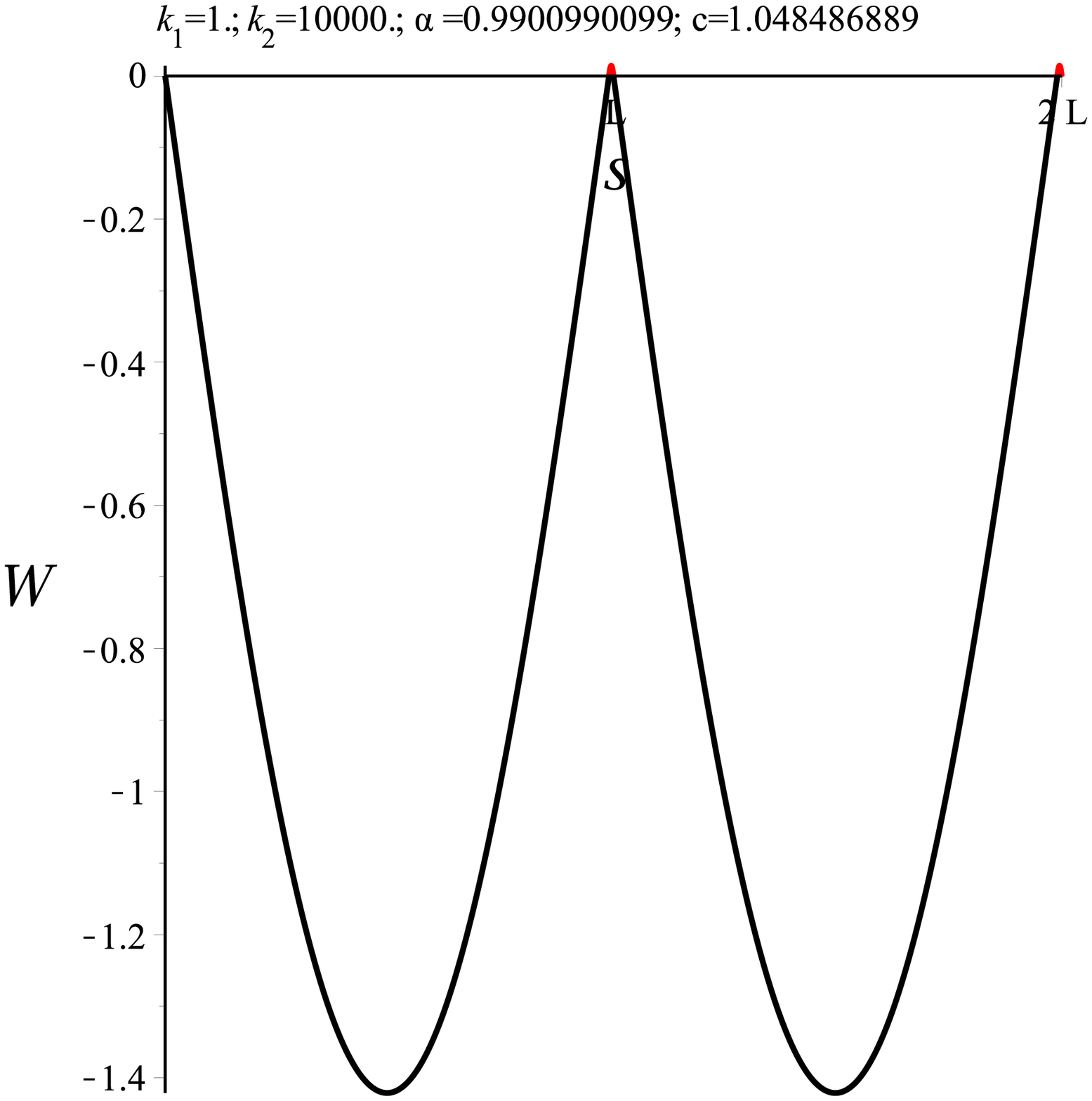}
b)  \includegraphics[width=6cm]{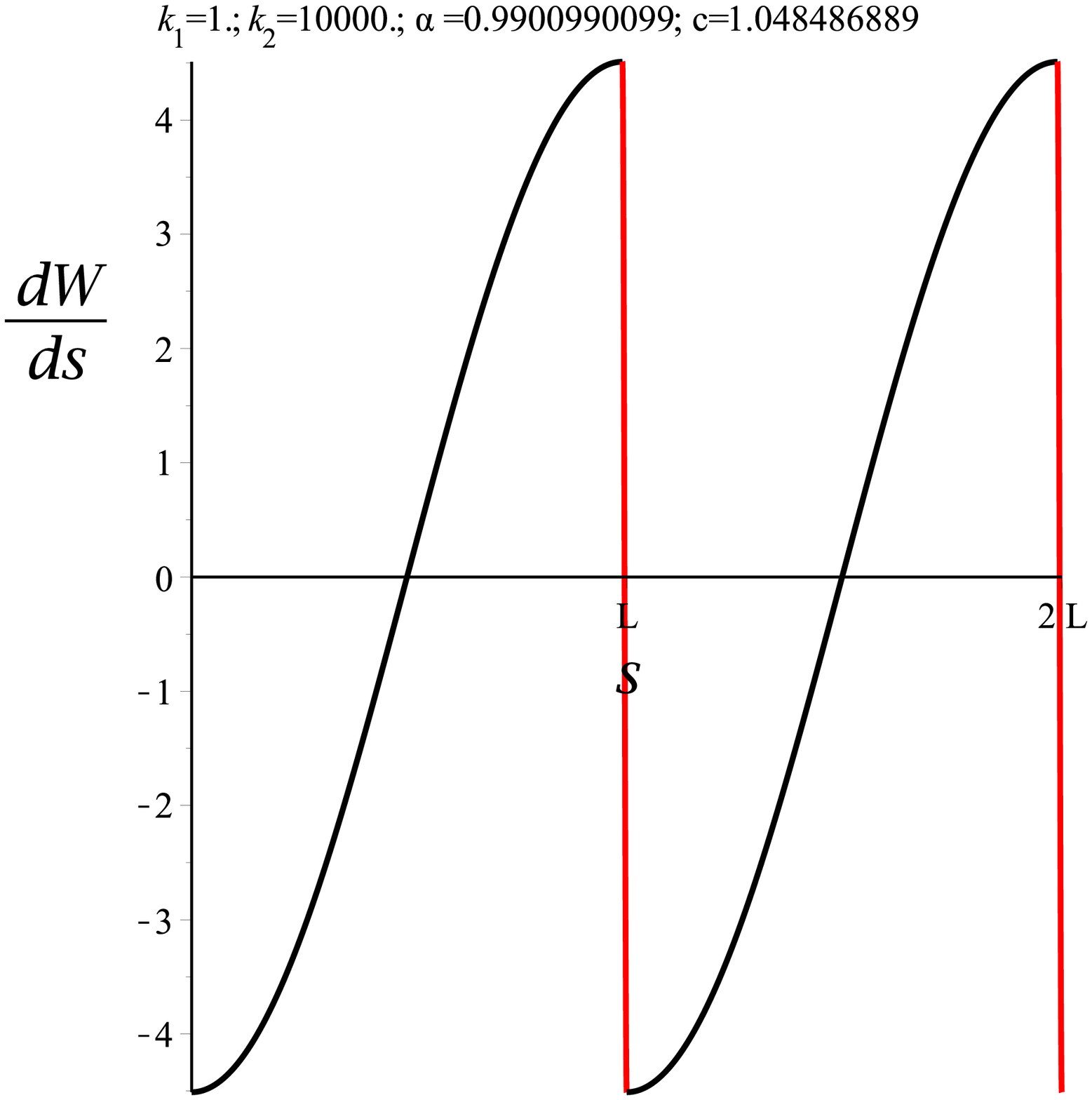}
  \caption{The solution $W(s)$ for $k_1=1$ and $k_2=10^{4}$. $n=1$.}
  \label{fig03}
\end{figure}

When $k_2 \rightarrow \infty$,  one side of the substrate becomes rigid. In this limit $\alpha \rightarrow 1$, $c \rightarrow 1+k_1/\pi^2$ and $W_2(\xi) \rightarrow 0$ (see (\ref{alpha}), (\ref{phasec}), (\ref{eq13a}) and (\ref{eq13b})), which means that the detached tension region reduces to one point ($s=L$) and a non-vanishing solution exists only on the side of the deformable substrate. An example is reported in Fig. \ref{fig03}a. Also in this case the derivative  has a jump (Fig. \ref{fig03}b) since

\begin{align}
\lim_{k_2 \rightarrow \infty} W_2'(\xi) \approx \lim_{k_2 \rightarrow \infty} \cos \left( \frac{\pi \xi n \sqrt{k_2}}{\sqrt{k_1}} \right) \label{eq18}
\end{align}
is undefined, although bounded as $k_2 \rightarrow \infty$. \\

The conclusion is that periodic waves with regular profile on unilaterally rigid substrate do not propagate in taut cables. \\

In this case, as it results from equations (\ref{eq06a})-(\ref{eq06b}) and at variance with the situation of section \ref{sec:unilaterallyrigid}, 
no totally confined solution exists in the in-contact compression interval ($w<0$ with strict inequality) at all.

\subsubsection{Transverse distributed load}\label{ssec:load}
When a fixed, constant transverse distributed load $\hat{p} = {\Tilde{p}/ \rho} \in \mathbb{R}$ is present equation (\ref{waveeq}) becomes
\beq
\frac{\partial^2 w}{\partial t^2}- v^2 \, \frac{\partial^2 w}{\partial x^2}+ \gamma(w) \, w =\hat{p}. \label{waveeqload}
\eeq
Also in this case, we look for solutions of the form
 (\ref{eq02}) with the substrate stiffness given by equations (\ref{stiffness1})-(\ref{stiffness2}). By using the dimensionless variables introduced in (\ref{eq03}), the equations for $W_1$ and $W_2$ (analogous to equations (\ref{eq04a}) and (\ref{eq04b})) are
 \begin{align}
& (c^2-1) \frac{d^2 W_1}{d \xi^2}(\xi) + k_1 W_1(\xi)=p, \quad \mathrm{for} \quad W_1(\xi) \leq 0, \label{eq04aload} \\
& (c^2-1) \frac{d^2 W_2}{d \xi^2}(\xi) + k_2 W_2(\xi)=p, \quad \mathrm{for} \quad W_2(\xi) \geq 0, \label{eq04bload}
\end{align}
where $p=\hat{p} \, \rho \, L^2 / T=\hat{p} L^2/v^2=\Tilde{p} L^2/T$ and the solution is given by
\begin{align}
&W_1(\xi)=c_1 \sin \left( a \xi \right) + c_2 \cos \left( a \xi \right) + \frac{p}{k_1}, \label{eq21a} \\
&W_2(\xi)=c_3 \sin \left( b  (\xi-\alpha) \right) + c_4 \cos \left( b (\xi-\alpha) \right) + \frac{p}{k_2}. \label{eq21b}
\end{align}
Note that equations (\ref{eq21a})-(\ref{eq21b}) support traveling wave solutions with arbitrary phase velocity and which do not change sign on the periodicity interval. This occurs when $\sqrt{c_1^2+c_2^2}<|p/k_1|$ 
with $p<0$ or $\sqrt{c_3^2+c_4^2}<|p/k_2|$ 
with $p>0$, i.e. if the oscillating part is 
smaller that the constant one.
The solution in this case remains positive or negative, according to the sign of $p$, on the whole interval. The coefficients $a$ and $b$ (respectively) remain undetermined; according to the nomenclature introduced in Section \ref{ssec:periodic} we call them ``zero" wave solutions (because there are no points at which the solution vanishes). \\

The symmetry property of the periodic travelling solutions without the transverse external load, discussed in Section \ref{ssec:periodic} immediately after equation (\ref{stiffness2}), here involves the external load $p$ as well: if $W(\xi)$ is a solution corresponding to a propagation velocity $c$ for given values of the stiffnesses $k_1$ and $k_2$ and the external load $p$, then $-W(\xi)$ is the solution with the same propagating speed $c$ and the stiffnesses reversed ($k_1 \to k_2$, $k_2 \to k_1$) and with $p \to -p$. \\

By imposing  the boundary conditions (\ref{eq07a})-(\ref{eq07d}) on the displacements we obtain the solutions which change sign on the periodicity interval; we have for the constants $c_1$, $c_2$, $c_3$ and $c_4$:
\begin{align}
&c_1=\frac{p}{k_1} \frac{\cos(a \alpha)-1}{\sin(a \alpha)}, \quad c_2=-\frac{p}{k_1}, \nonumber \\
&c_3=\frac{p}{k_2} \frac{\cos\left[b \left(1 - \alpha\right) \right]-1}{\sin\left[b \left(1 - \alpha\right) \right]}, \quad c_4=-\frac{p}{k_2}. \label{eq22}
\end{align}
It is easy to see, with some algebra, that
\beqn
&& c_1^2+c_2^2 = \left(\frac{p}{k_1}\right)^2 \, \sec^2 \left(\frac{a \, \alpha}{2} \right) \ge \left(\frac{p}{k_1}\right)^2 \\
&& c_3^2+c_4^2 = \left(\frac{p}{k_2}\right)^2 \, \sec^2 \left(\frac{b \, (\alpha-1)}{2} \right) \ge \left(\frac{p}{k_2}\right)^2
\eeqn
in agreement with the request that the solutions change sign over the period. Note also that they exist if and only if $a \alpha \neq \pi$ and $b \left(1 - \alpha\right) \neq \pi$, which are complementary to the existence conditions  (\ref{eq07e}) of the case without transverse load, which then comes out to be a special ``resonant" case. An additional condition is given by the request that $W_1(\xi) \le 0$ for $0 \le \xi \le \alpha$ and $W_2(\xi) \ge 0$ for $\alpha \le \xi \le 1$, for which it is sufficient that $W_1'(0) \le 0$. By using equations (\ref{eq21a}) and (\ref{eq22}) we obtain $W_1'(0) = a \, c_1 \le 0$ which entails $c_1 \le 0$, that is
\beq
\frac{p}{k_1} \frac{\cos(a \alpha)-1}{\sin(a \alpha)} \le 0.  \label{condminmax}
\eeq
This condition indicates that for $a \alpha  \le \pi$ we must have $p>0$, while for $a \alpha  \ge \pi$ we must have $p<0$. \\

The two boundary conditions (\ref{eq11}) and (\ref{eq12}) on the derivatives are again linearly dependent (for the same geometrical reasons), and give, after some algebra,
\begin{align}
  & \sqrt{\frac{k_1}{k_2}} = \frac{\sin(a \alpha)}{\cos(a \alpha)+1} \frac{\sin\left[a \sqrt{\frac{k_2}{k_1}} \left(1 - \alpha\right) \right]}{\cos\left[a \sqrt{\frac{k_2}{k_1}} \left(1 - \alpha\right) \right]-1}, \label{eq23}
\end{align}
which  provides a relationship between $a$ and $\alpha$ in terms of the ratio $k_1/k_2$, e.g., $a=a(\alpha,k_1/k_2)$. With respect to the case without load, where for given values of $k_1$ and $k_2$ the parameters $\alpha$ and $c$ are uniquely determined, here, instead, $\alpha$ can be chosen freely and $a$ is then determined as the solution of the trascendental equation (\ref{eq23}). Thus, we have a 1-parameter manifold of  solutions. The value of $\alpha$ for which $a \, \alpha = \pi$, which is obtained when $\alpha$, $c$ and $a$ are given by equations (\ref{eq05}), (\ref{alpha}) and (\ref{phasec}), i.e. their expressions in the case without load, is crucial for the fulfillment of condition (\ref{condminmax}) and we shall call it $\alpha_{cr}$ (critical value): for $\alpha = \alpha_{cr}$ the solution of equations (\ref{eq21a})-(\ref{eq23}) ceases to exist and the amplitude of the wave, which is no longer a free parameter as in the absence of the load, but it depends on $\alpha$, diverges in the limit; for $\alpha \le \alpha_{cr}$ the solution exists for $p>0$, while for $\alpha \ge \alpha_{cr}$ the solution exists for $p<0$. We shall illustrate this fact by numerical examples in Section \ref{sec:numerical}. \\

\begin{figure}[h!]
  \centering
    \includegraphics[width=7cm]{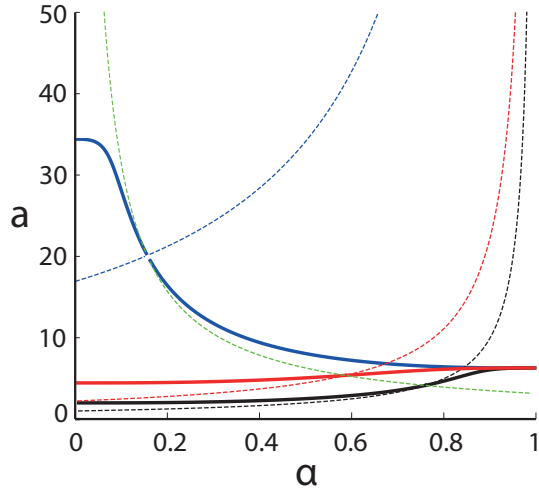}
  \caption{The solutions of equation (\ref{eq23}) for $k_1=30,k_2=1$ (solid red line), $k_1=1,k_2=2$ (solid black line) and $k_1=1,k_2=10$ (solid blue line). The green dashed line represents the curve $a=\pi/\alpha$ corresponding to the case without load. The other dashed lines represent the second of (\ref{eq07e}), $b \, (1-\alpha) = \pi$.}
  \label{fig04}
\end{figure}

For a given $\alpha$, $a$ is determined by equation (\ref{eq23}) and the propagation speed of the wave is then given by
\beq
c=\sqrt{1+\frac{k_1}{a^2}}.\label{cload}
\eeq
 The solutions of equation (\ref{eq23}) are reported in Fig. \ref{fig04} as curves in the cartesian plane $(\alpha,a)$. The solid lines represent the solutions for $k_1=30$ and $k_2=1$ (red line), $k_1=3$ and $k_2=1$ (black line) and $k_1=0.1$ and $k_2=1$ (blue line). The dashed green line represents the first of the existence conditions (\ref{eq07e}), $a \, \alpha = \pi$ of the solution without load, while the other dashed lines represent the second of (\ref{eq07e}), $b \, (1-\alpha) = \pi$. The intersections of the dashed curves with the solid lines represent graphically the values of $\alpha_{cr}$ for the corresponding values of $k_1$ and $k_2$. Note that $a(0,k_1/k_2)=2 \, \pi \, \sqrt{k_1/k_2}$ and $a(1,k_1/k_2)=2 \, \pi$.\\

\section{Numerical Simulations} \label{sec:numerical}

This section is devoted to complement the analytic solutions found in Section \ref{cable} with some numerical investigations. A numerical solution of the wave equation (or, for that matter, of any evolution equation) can only be sought for in the form of a Cauchy problem, with a given set of initial and boundary conditions. Therefore, an exact reproduction of our periodic analytic solutions is not attainable; however, with a suitable choice of the initial and boundary conditions we can generate solutions which, after a relatively short transient, follow the pattern predicted by our analytic model in an interval of the whole real domain. The Cauchy problem for our simulations is given by equations (\ref{kgnl})-(\ref{kgnlic}) with the boundary condition $\vphi(t)$ equal to the analytical solution at $x=0$,
$$
\vphi(t) = w(0,t), \label{boundary}
$$
where $w(x,t)$ is the periodic solution given by (\ref{eq02}), (\ref{eq13a}), (\ref{eq13b}) and the initial condition is $w_0(x)=\psi(x)=0$.  In addition, we impose that the numerical solution $w(x,t)$ vanishes at the end of the simulation domain.  The undetermined amplitude (see equations (\ref{eq06a}), (\ref{eq06b}), (\ref{eq13a}) and (\ref{eq13b}) with the following discussions) is indicated by $a_0$ in the figure captions. With this choice, because of well-known properties of the wave equation \cite{john}, two symmetric waves are generated at the origin, a left-propagating wave and a right-propagating wave. The latter, after an initial transient, becomes closer and closer to the periodic wave obtained analytically in Section \ref{cable}, restricted to the half-space $x \ge 0$. We will use this scheme as verification of the properties of our analytic solutions and also as a mean to assess the stability of our waves, both with and without external distributed load. The algorithm is a simple  finite-difference forward scheme, with the care of choosing the spatial discretisation and the time step properly in order to avoid numerical instabilities. Equation (\ref{dEdt}) has been used as a benchmark in all applicable cases and it is fulfilled with an accuracy of $10^{-4}$ to  $10^{-5}$.

\subsection{No transverse distributed load}
\begin{figure}[h!]
  \centering
  \includegraphics[width=3.2cm]{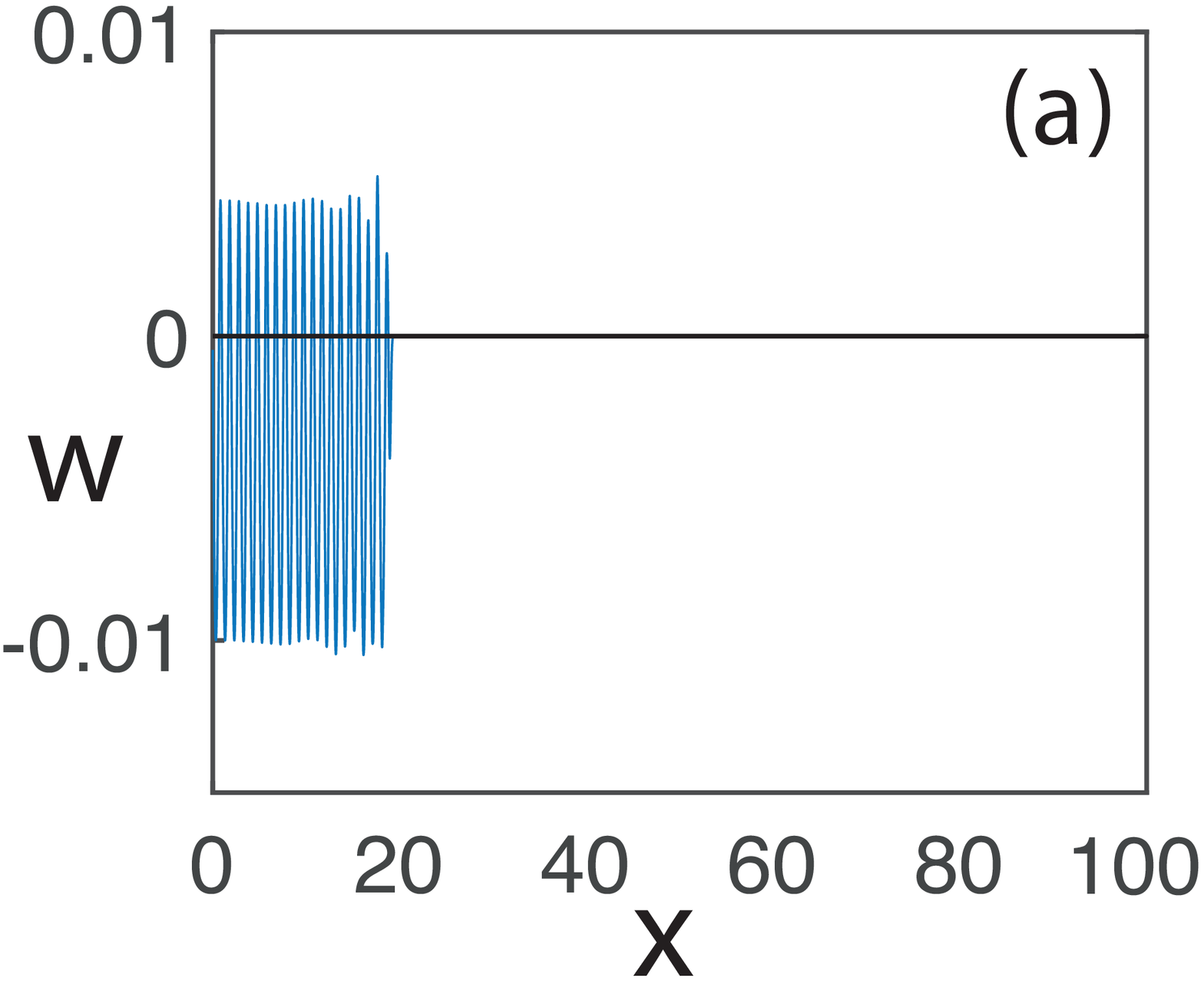}
  \includegraphics[width=3.2cm]{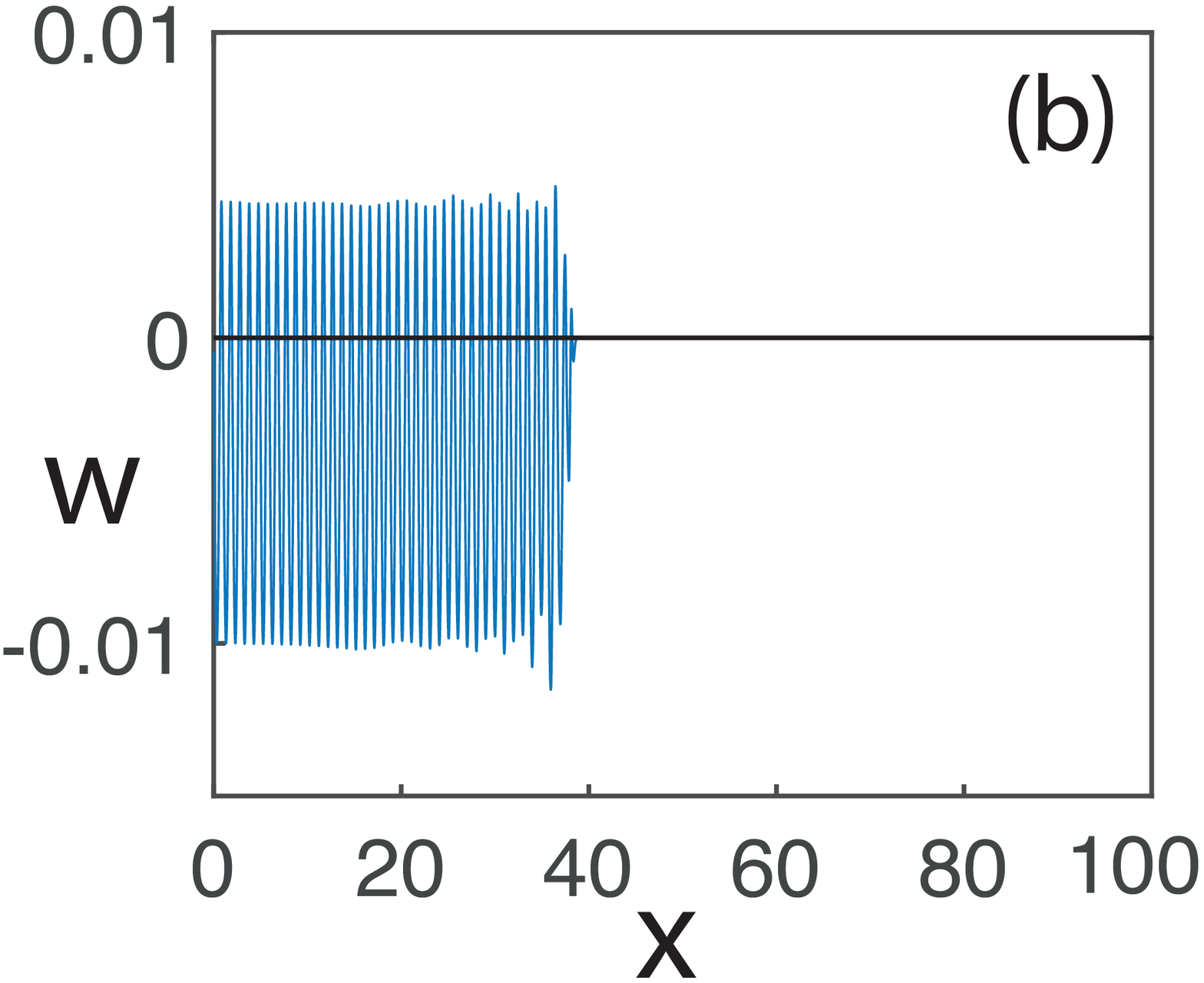}
  \includegraphics[width=3.2cm]{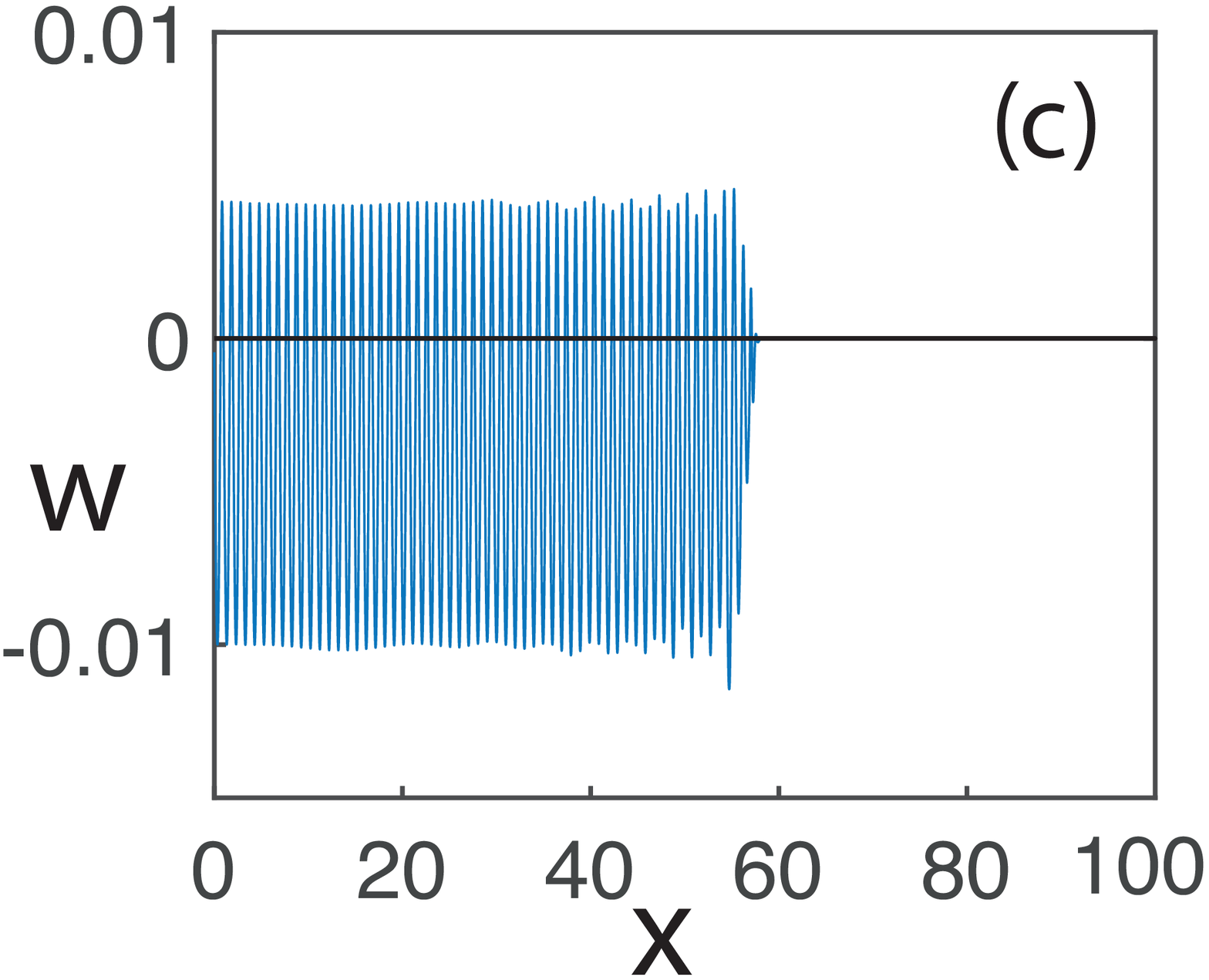}
  \includegraphics[width=3.2cm]{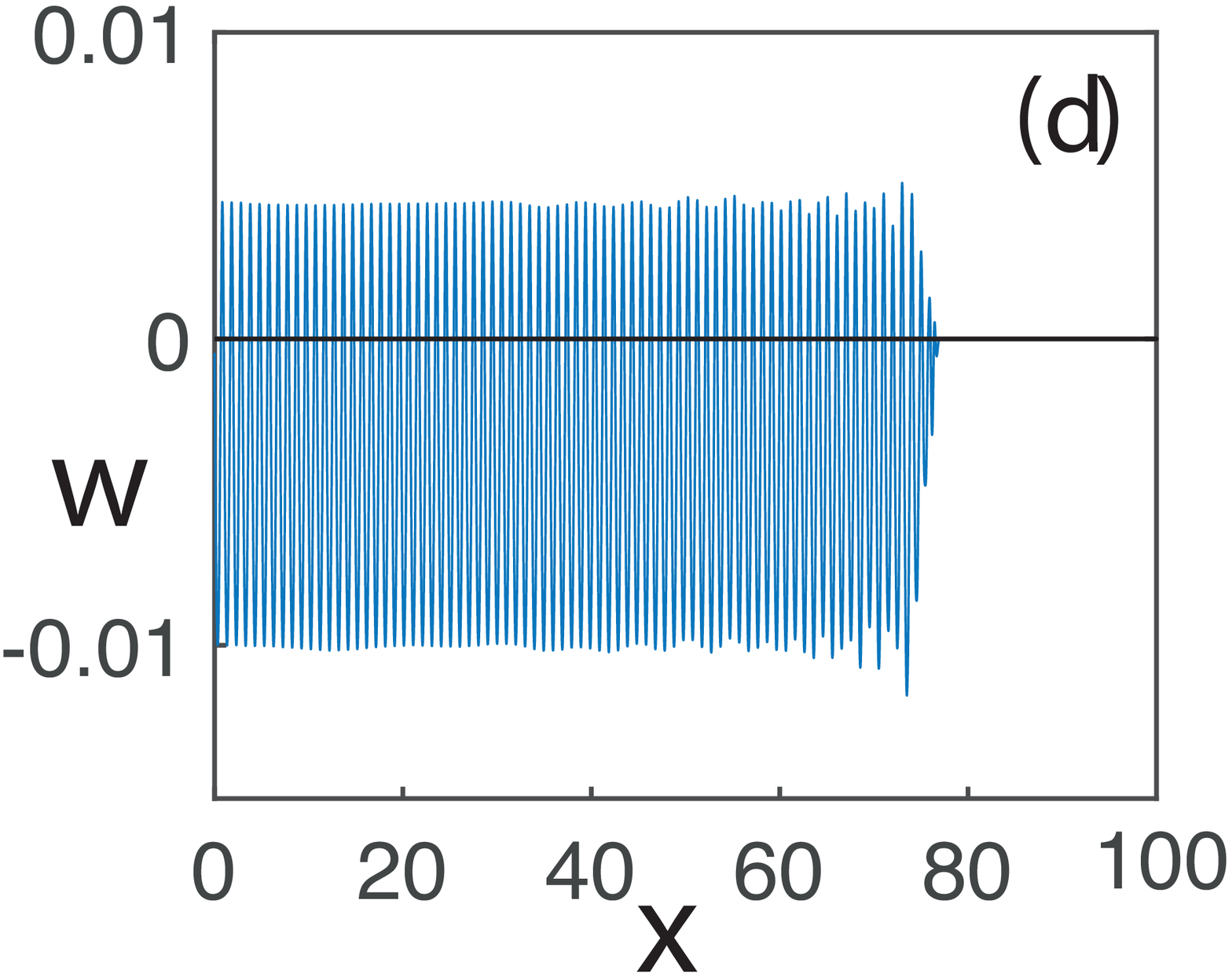}
   \caption{The solution $w(x,t)$ at four times for $k_1=1,k_2=5$ and $L=1$, $a_0=0.01$: (a) $t=20$, (b) $t=40$, (c) $t=60$, (d) $t=80$.}
  \label{fig101b}
\end{figure}
 
\begin{figure}[h!]
  \centering
 \includegraphics[width=6cm]{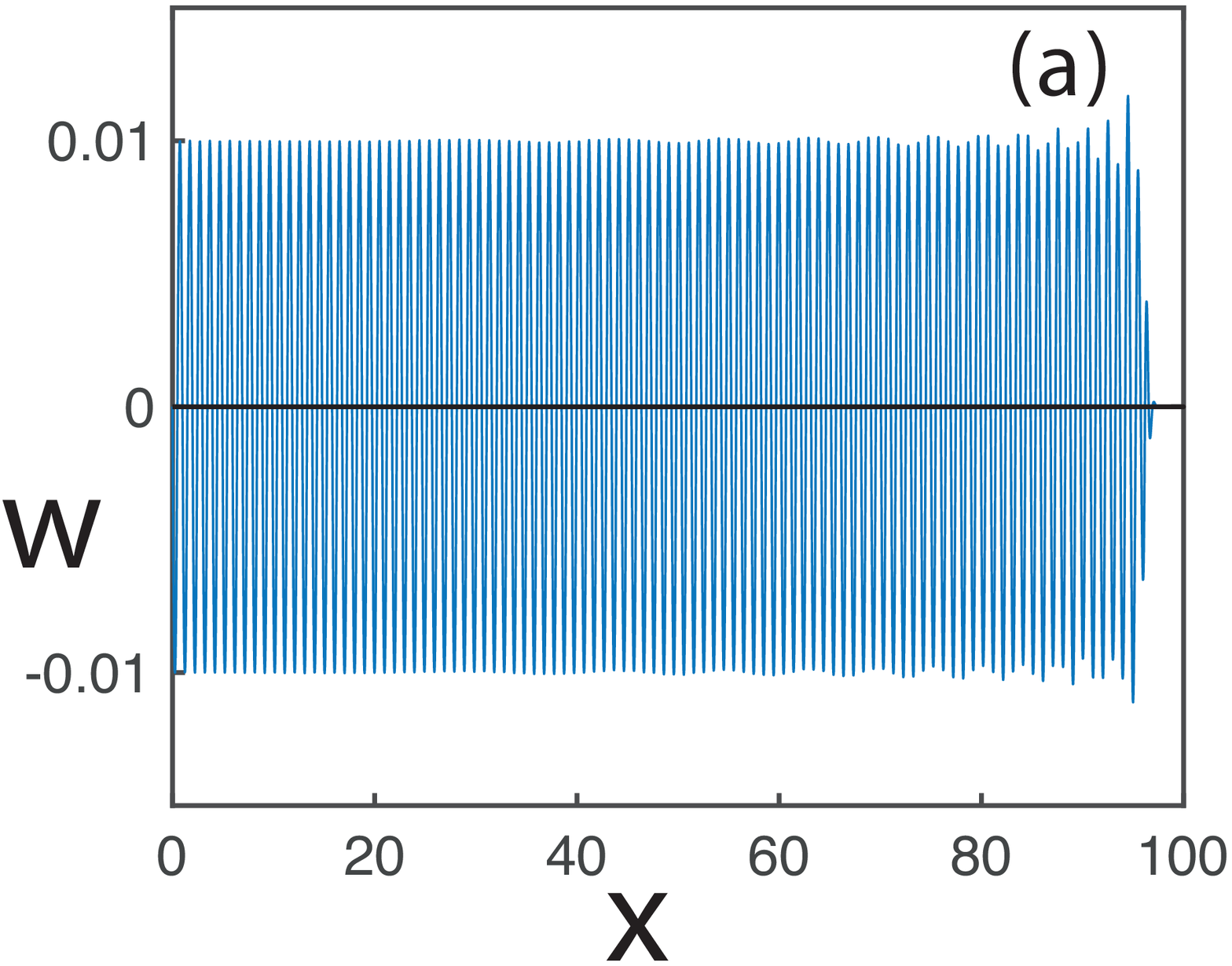}
  \includegraphics[width=6cm]{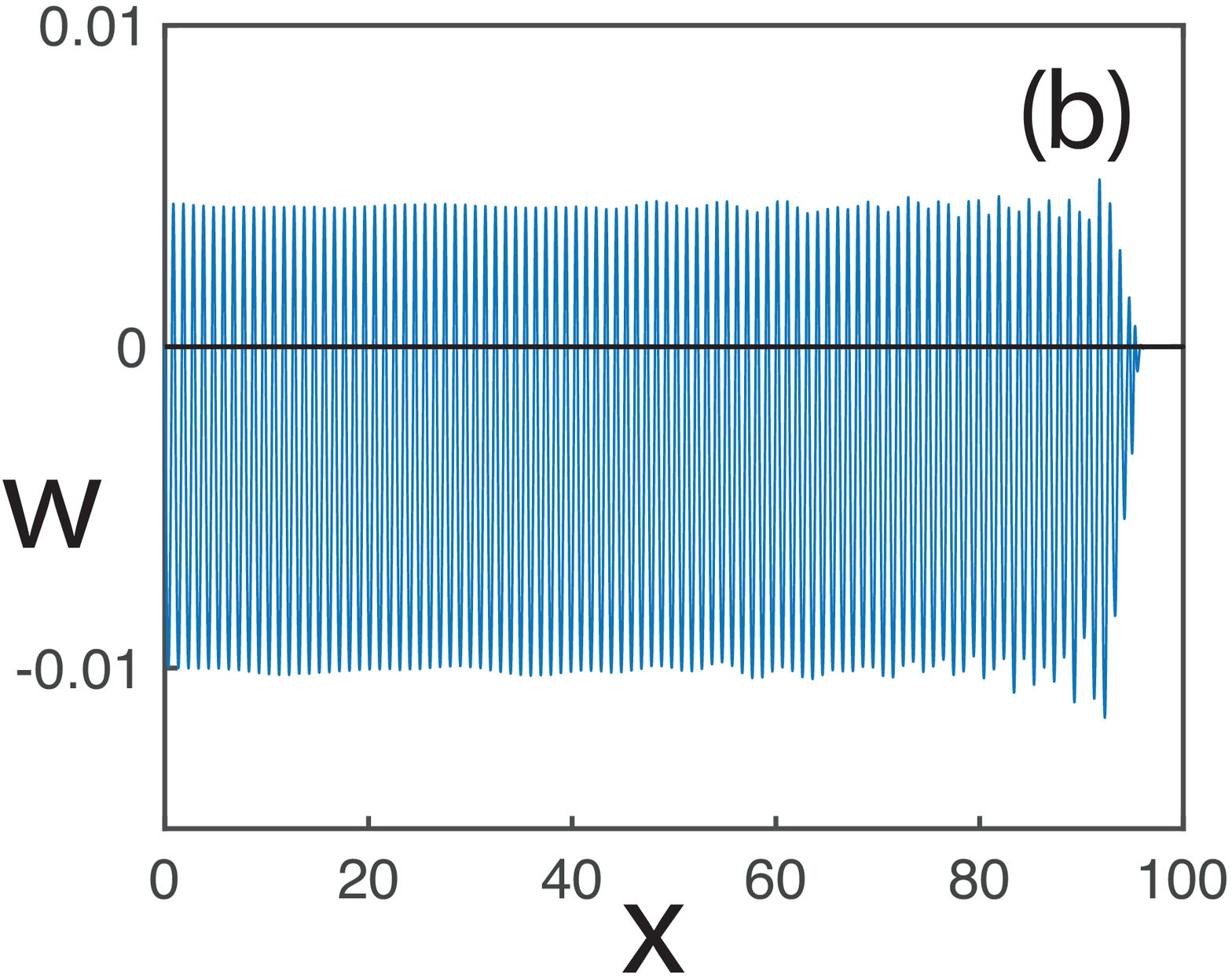}
 \hspace*{0.2cm}   \includegraphics[width=6.3cm]{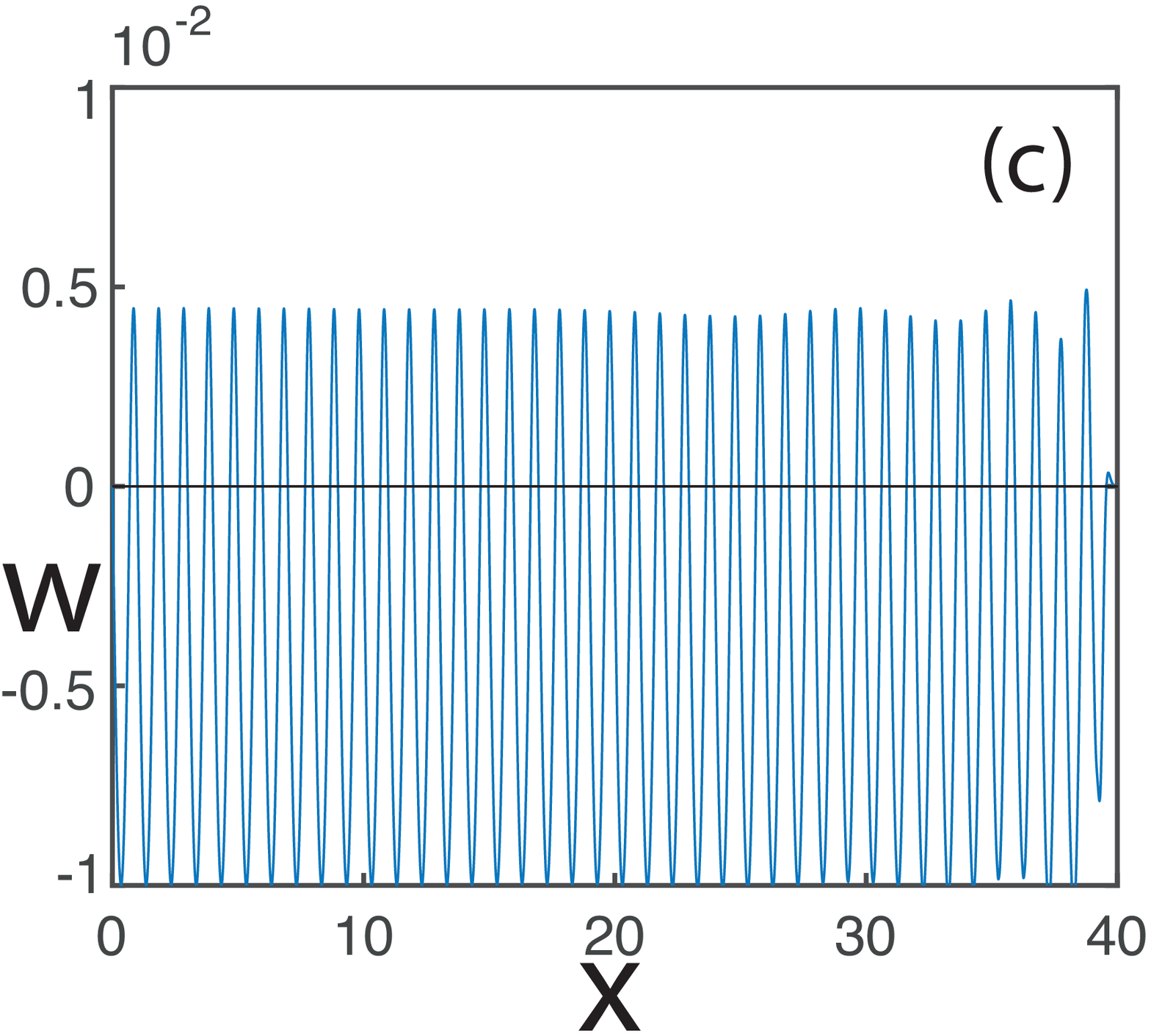}
  \includegraphics[width=5.6cm]{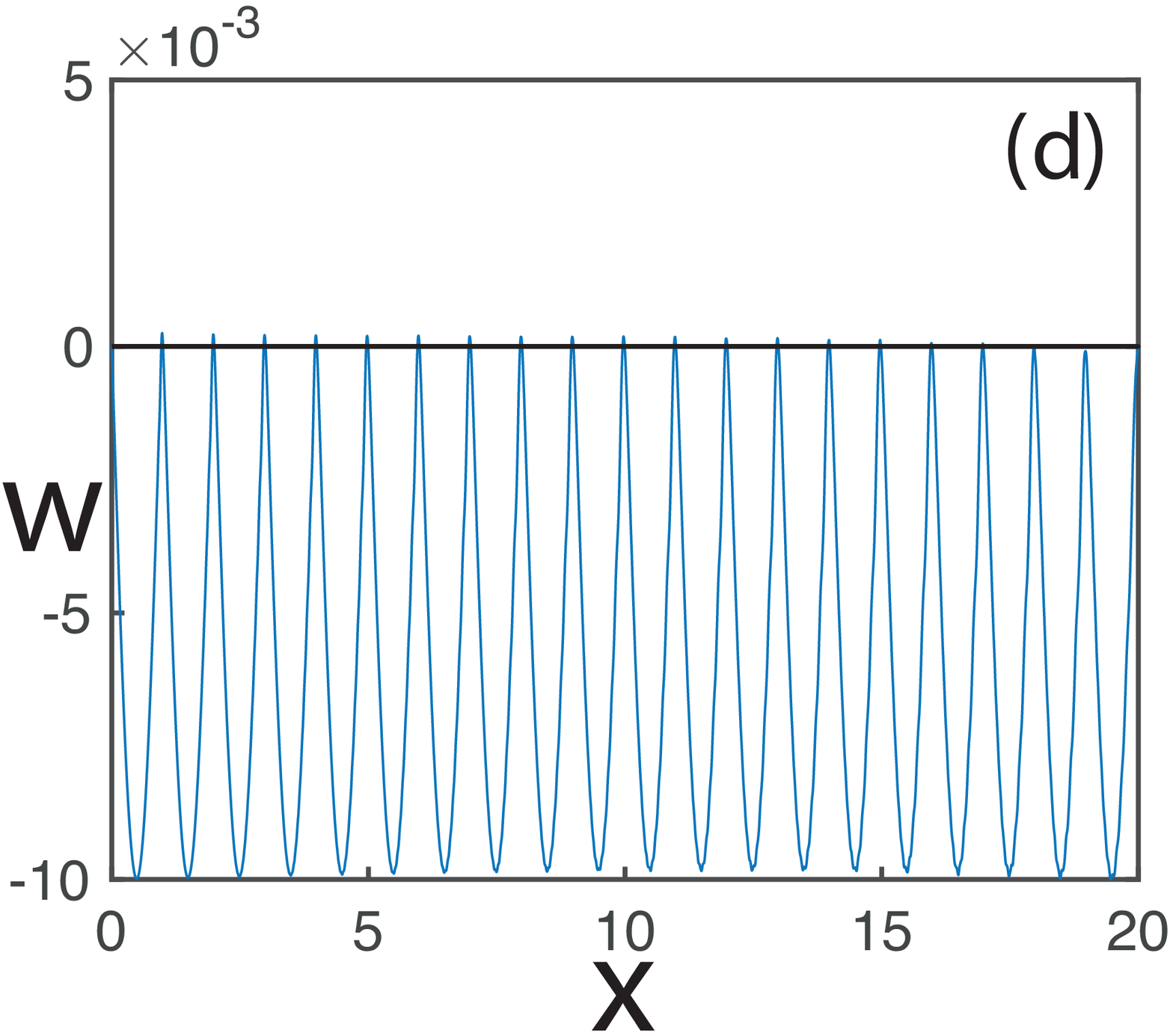}
  \caption{The solution $w(x,t)$ at the final time for (a) $k_1=k_2=1$ {\corrL (linear case)}, (b) $k_1=1,k_2=5$, (c) $k_1=0.2,k_2=1$, (d) $k_1=0.001, k_2=1$ and $L=1$, $a_0=0.01$.}
  \label{fig101}
\end{figure}

The solution $w(x,t)$ without the external load is shown in Figures \ref{fig101b}(a-d) and \ref{fig101}(a-d). Figures \ref{fig101b}(a-d) show the progression of the periodic traveling wave with snapshots at constant time intervals for $L=1$, $a_0=0.01$ and $k_1=1,k_2=5$; they illustrate how the wave generated by the boundary condition at the origin propagates into the initially empty half-space $x \ge 0$ with a phase speed consistent (within numerical errors) with the speed given by equation (\ref{phasec}). An initial transient is formed on the head of the wave front and propagates to the right, while in an increasingly larger portion of the domain the solution settles to a right-propagating wave with constant amplitude, which is the restriction to $x \ge 0$ of the wave found analytically in Section \ref{cable} (the wave profile at the final time is shown in figure \ref{fig101}(b)). Figures \ref{fig101}(a-d) show the solution at the final time of our simulations for $L=1$, $a_0=0.01$ and (a) $k_1=k_2=1$, (b) $k_1=1,k_2=5$, (c) $k_1=0.2,k_2=1$, (d) $k_1=0.001, k_2=1$. The solution portrayed in Figure \ref{fig101}(a) corresponds to the bilateral case; the maximum excursions in Figures \ref{fig101}(b,c), both for $w > 0$ and $w < 0$ correspond to the excursions extracted analytically from equations (\ref{eq13a}) and (\ref{eq13b}). The solution shown in Figure \ref{fig101}(d), with a very small value of $k_1$, corresponds to the limit of Subsection \ref{sec:unilaterallyrigid}, with the proviso that, instead of $k_2 \to \infty$, we have chosen $k_1 \to 0$. We note that in this case, according to the theoretical results of section \ref{unilateral}, non differentiable points appear at the endpoints of the spatial period. In all cases covered by our simulations (including the many ones not shown in this paper) only the initial transient depends upon the stiffnesses $k_1$ and $k_2$  separately, while the amplitude and the frequency of the final right-propagating wave depends only upon the ratio $k_1/k_2$, in agreement with the theoretical model.
Finally, we remark that the choice of following the solution for a different number of periods in each case is due to numerical reasons, since the algorithm requires smaller time steps as $k_1 \to 0$ (or $k_2 \to \infty$), so we stop the simulation when the long-time behaviour is  evident. {\corrL The effect of the nonlinearity of the model when $k_1 \ne k_2$ is evident from the asymmetry of the wave profile with respect to the $w=0$ baseline; the effect becomes more pronounced in the limiting cases of unilateral and unilaterally rigid substrate}. The same pattern observed in Figures \ref{fig101}(a-d) is also observed on a wide range of values in the $(k_1,k_2)$ parameter space. \\

\begin{figure}[h!]
  \centering
  \includegraphics[width=4.3cm]{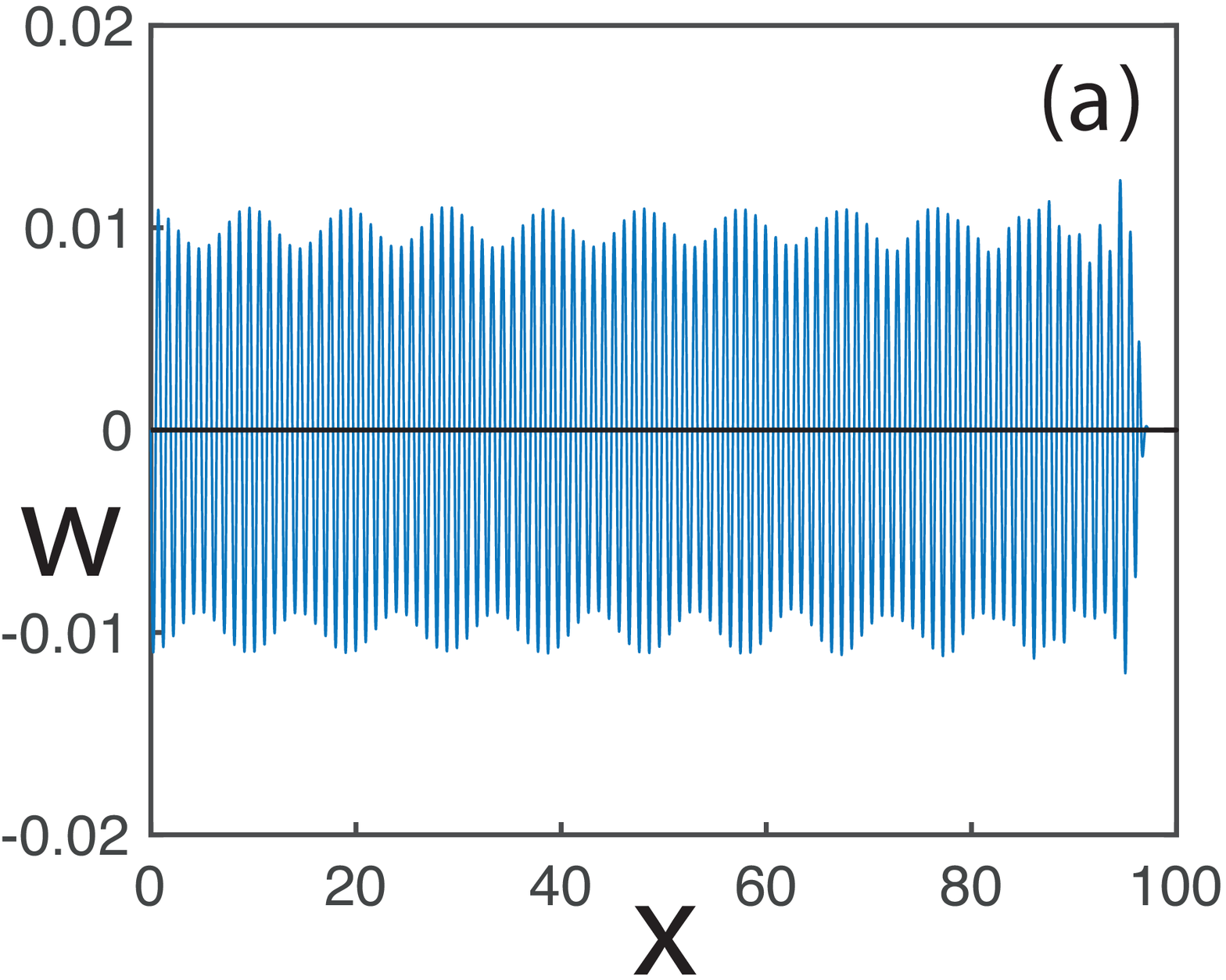}
  \includegraphics[width=4.3cm]{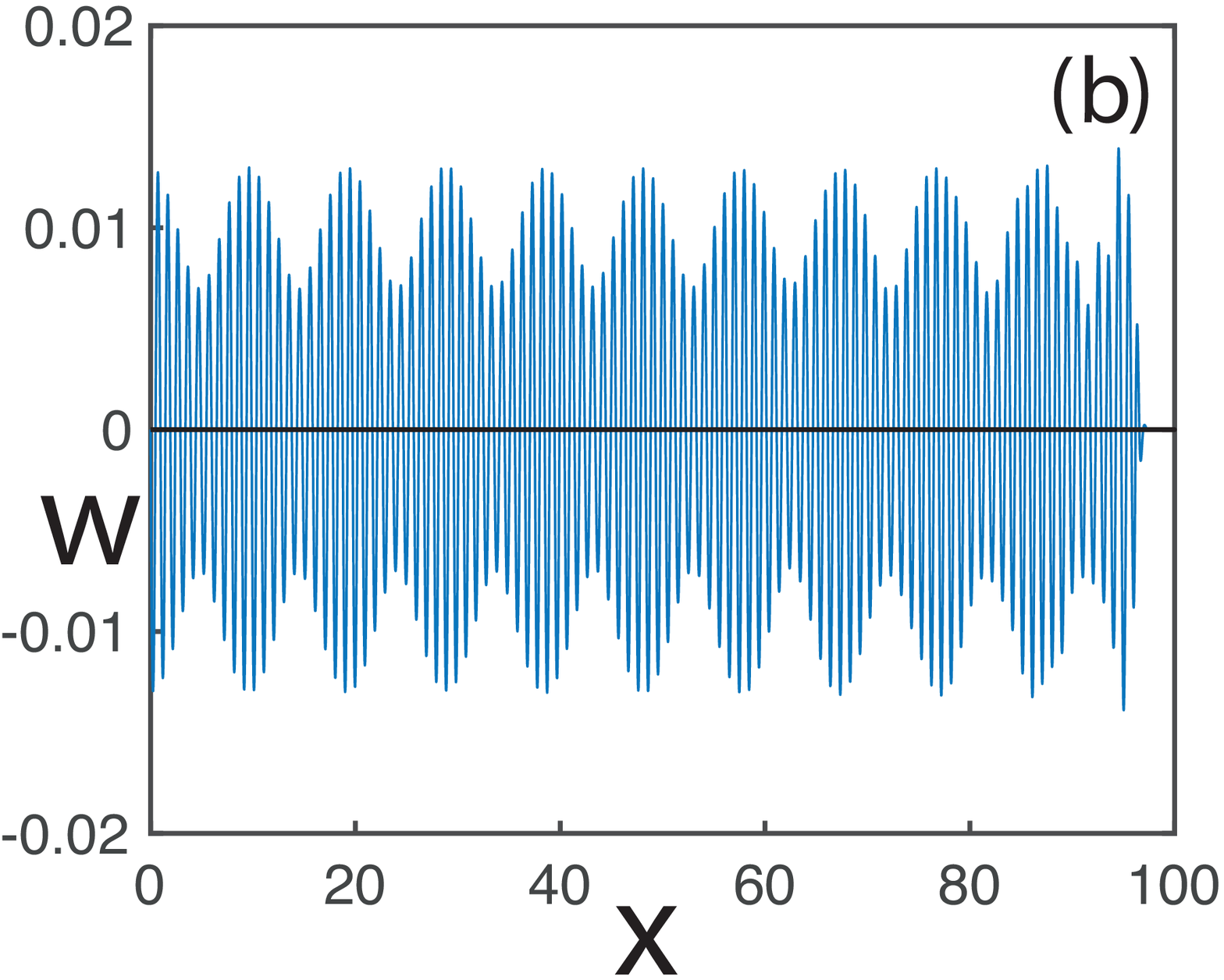}
  \includegraphics[width=4.3cm]{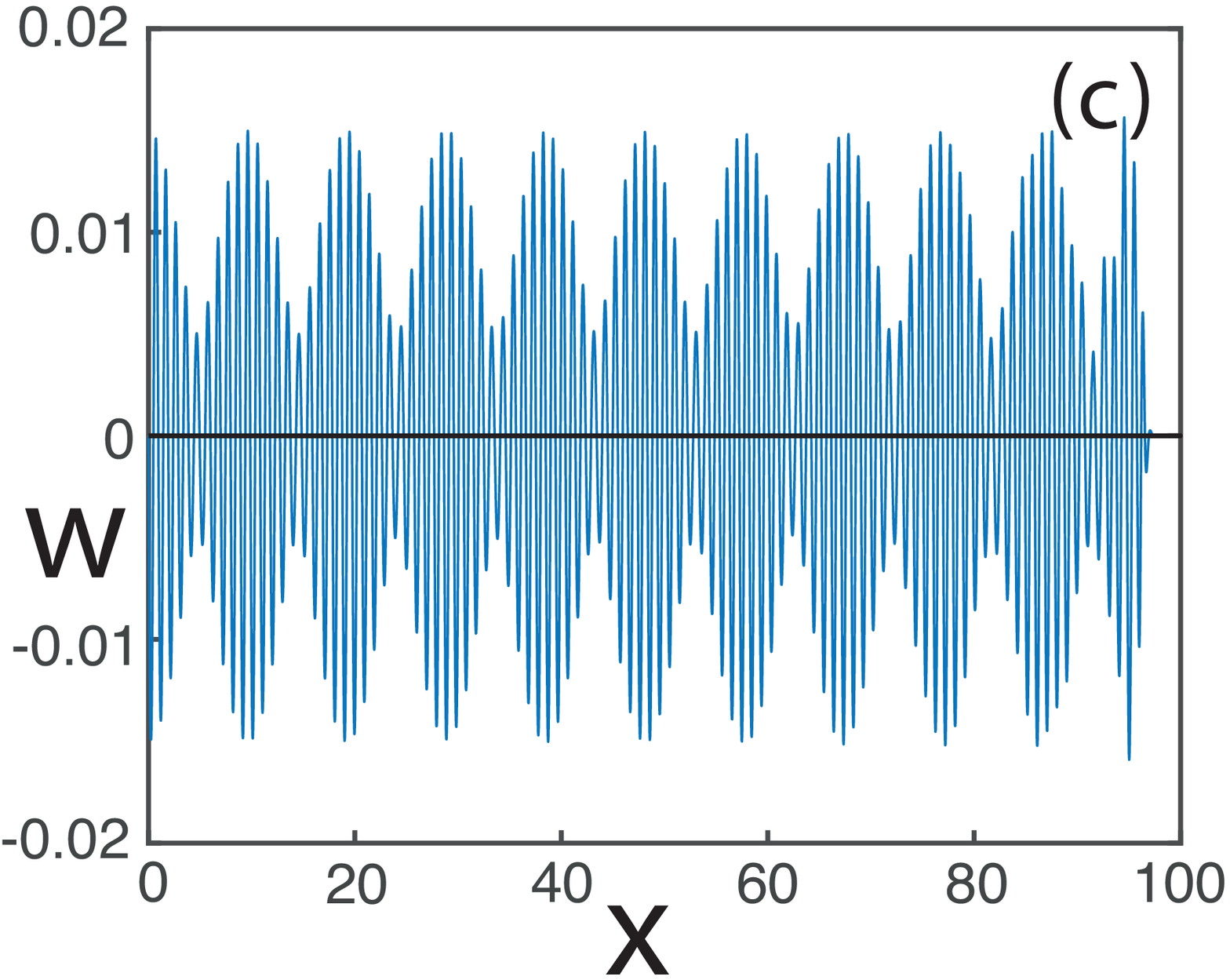}
  \caption{The solution $w(x,t)$ for $k_1=k_2=1$ {\corrL (linear case)}, with (a) $\veps=0.001$, (b) $\veps=0.003$, (c) $\veps=0.005$,  $L=1$, $a_0=0.01$ and $\omega_1=1.1 \, \omega$.}
  \label{fig102}
\end{figure}
\begin{figure}[h!]
  \centering
  \includegraphics[width=4.2cm]{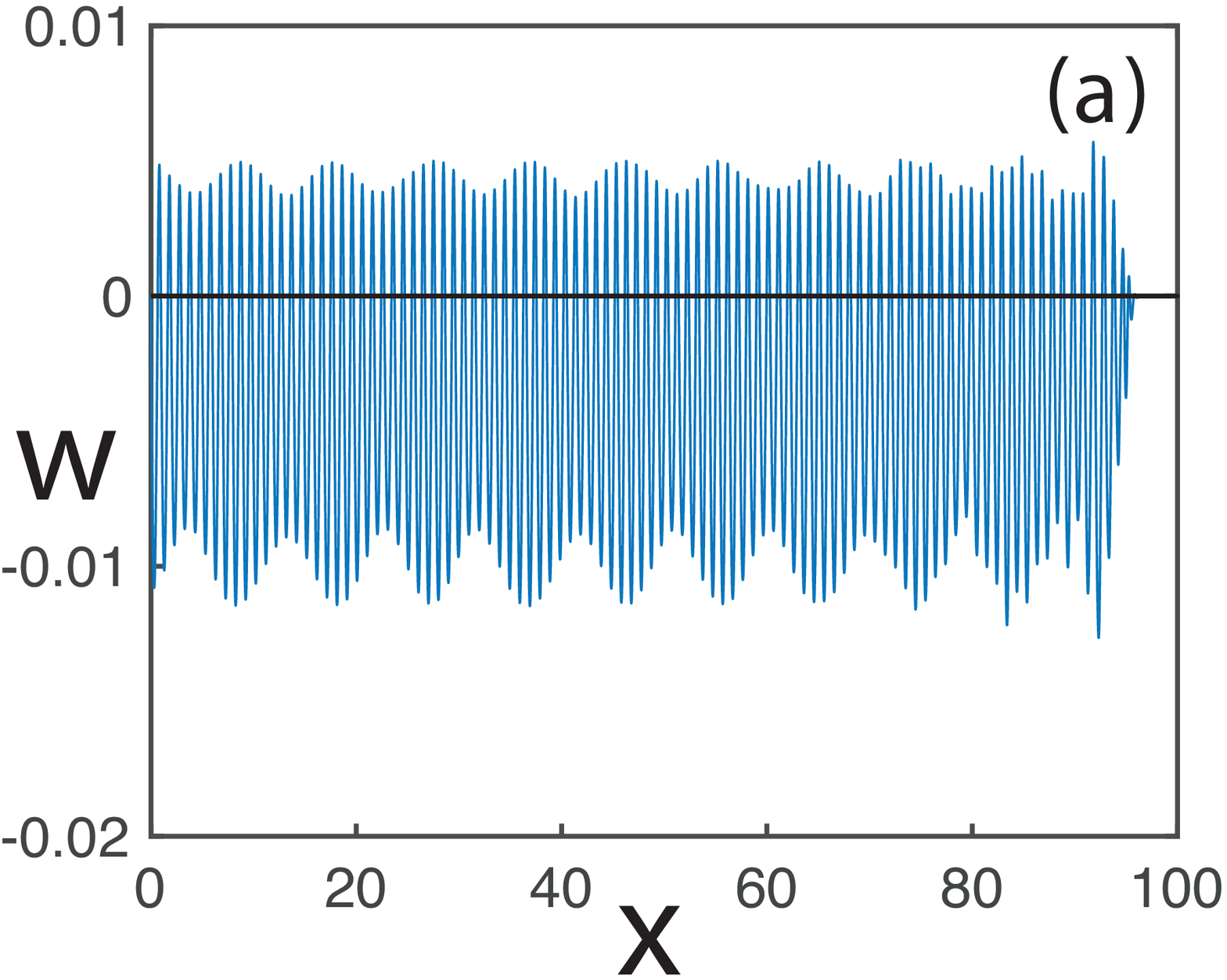} \hspace*{0.2cm}
  \includegraphics[width=4.2cm]{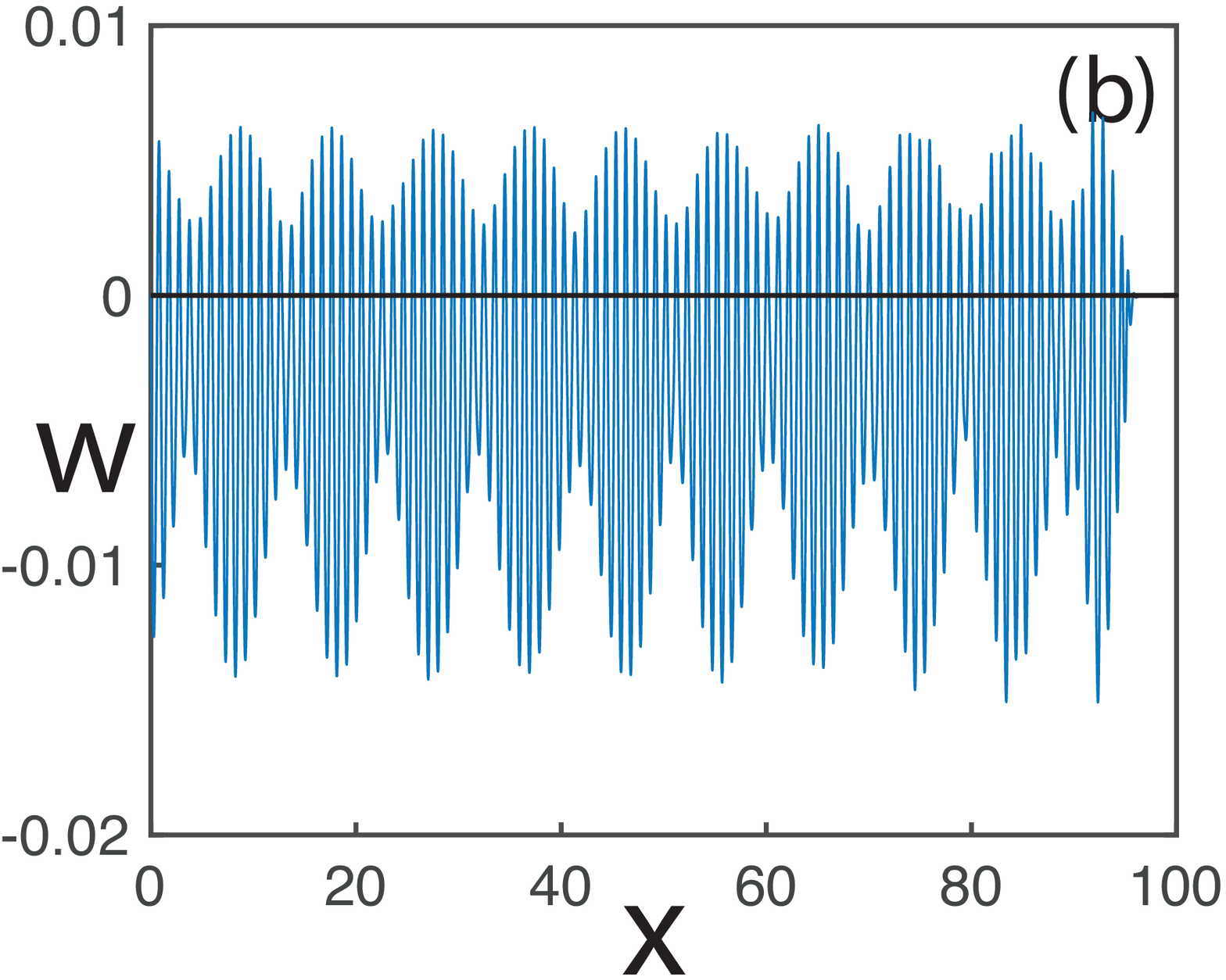}\hspace*{0.2cm}
  \includegraphics[width=4.2cm]{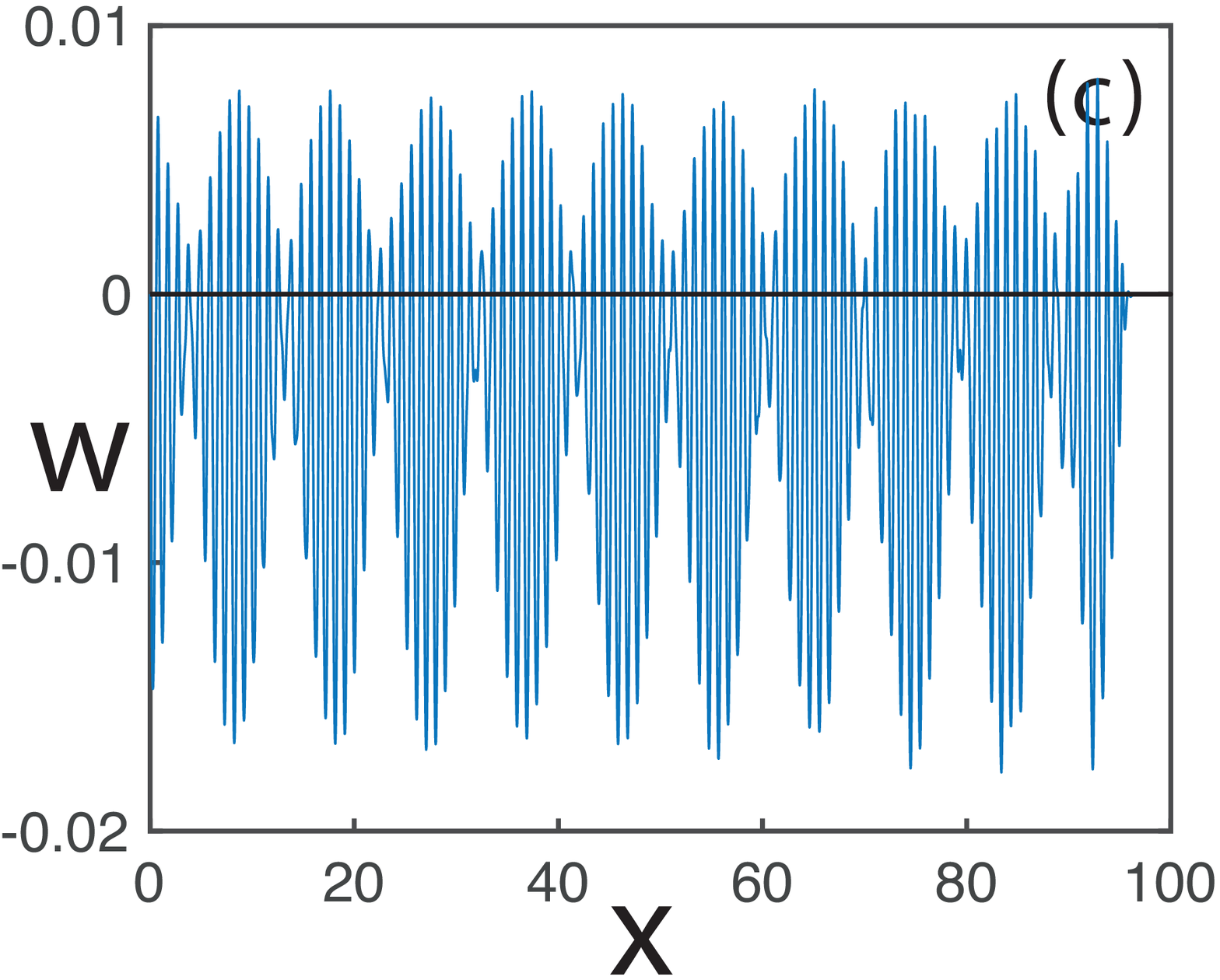}
  \caption{The solution $u(x,t)$ for $L=1$, $a_0=0.01$, $k_1=1,k_2=5$, $\omega_1=1.1 \, \omega$ and (a) $\veps=0.001$, (b) $\veps=0.003$,  (c) $\veps=0.005$.} \label{fig104}
\end{figure}
\begin{figure}[h!]
  \centering
  \includegraphics[width=4.4cm]{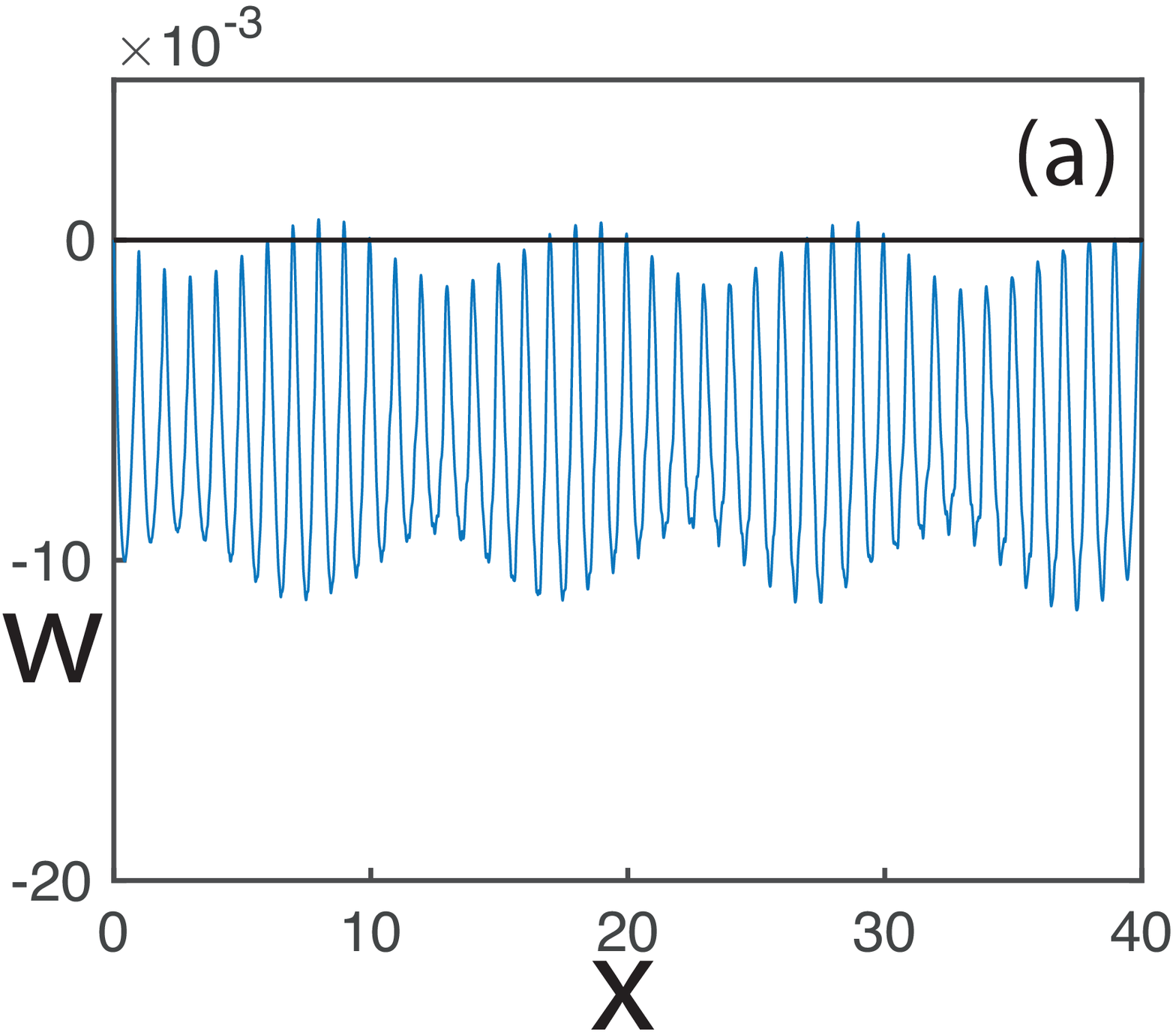} 
  \includegraphics[width=4.4cm]{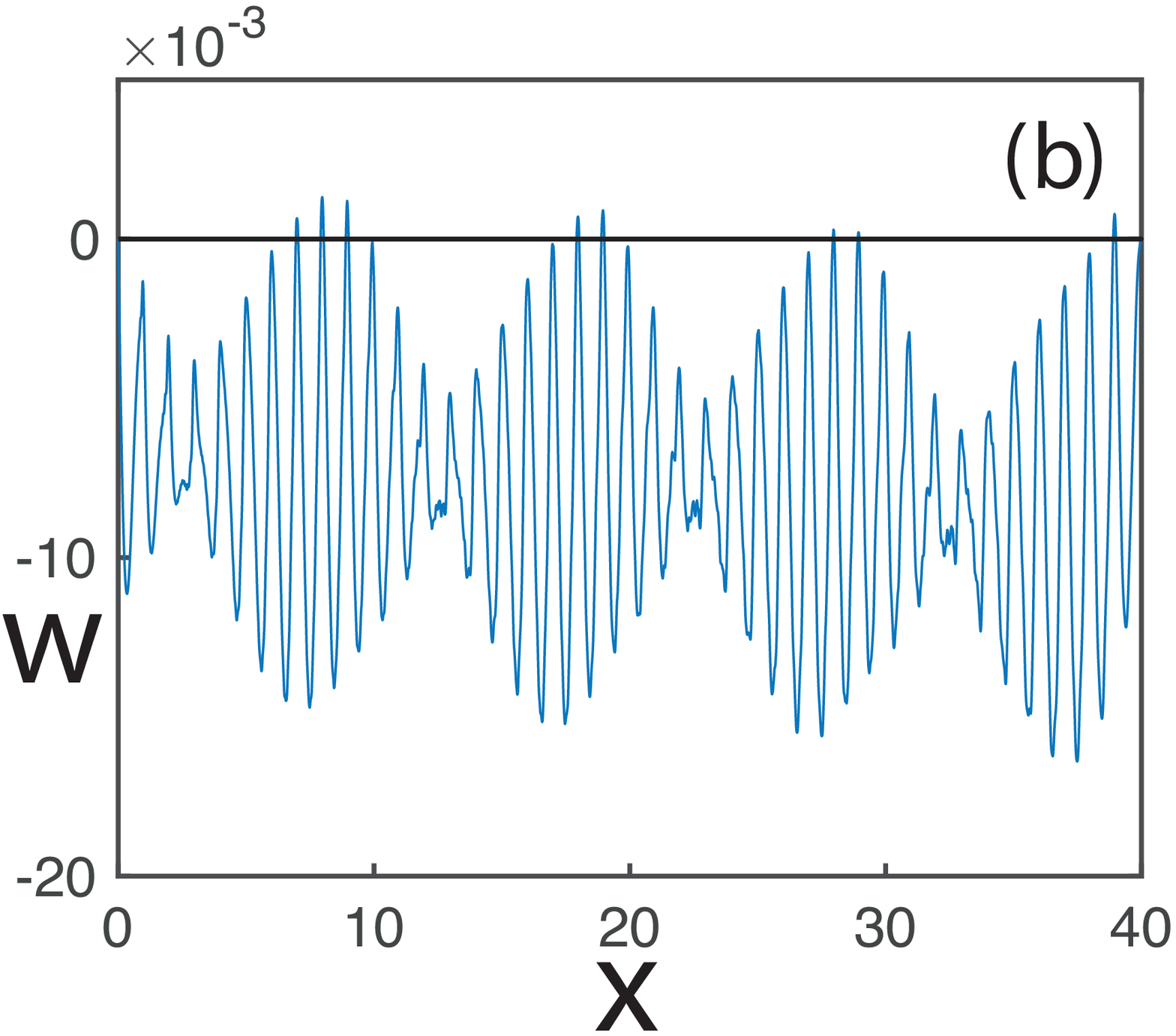} 
  \includegraphics[width=4.4cm]{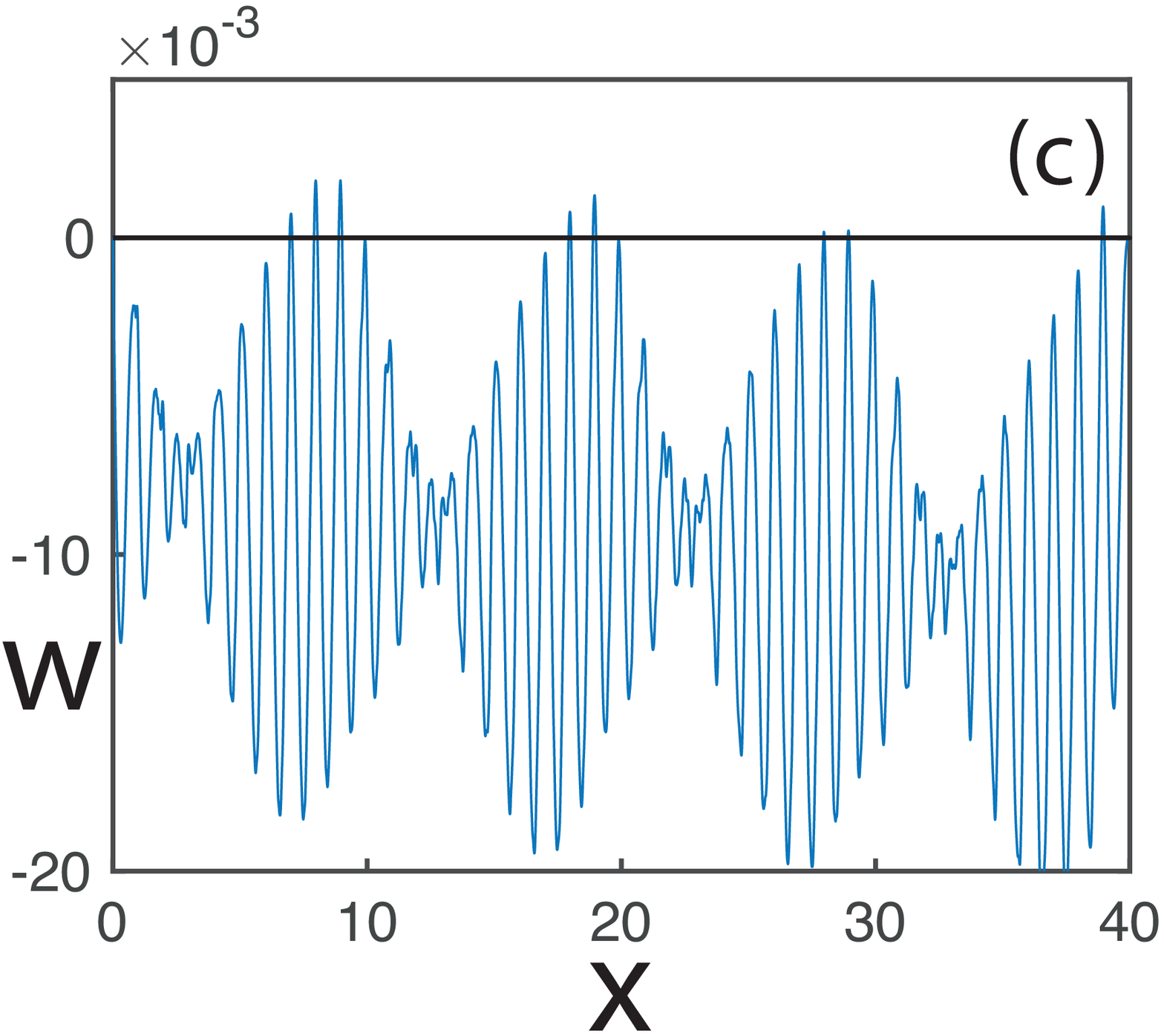}
  \caption{The solution $u(x,t)$ for $L=1$, $a_0=0.01$, $k_1=0.001, k_2=1$, $\omega_1=1.1 \, \omega$ and (a) $\veps=0.001$, (b) $\veps=0.003$,  (c) $\veps=0.005$.} \label{fig105}
\end{figure}

To address the stability properties of the periodic solutions obtained in the previous sections, we follow numerically solutions starting in the vicinity of the periodic traveling waves. We do so in two different ways: by imposing a small harmonic perturbation to the boundary condition or by choosing a small non-zero initial condition. With the first approach, let
\beq
\vphi(t) = w(0,t)+ \veps \, \sin (\omega_1 \, t).   \label{perturb}
\eeq
be the boundary condition, with $\psi(x)=0$. In Figure \ref{fig102}(a-c) we show the solution $w(x,t)$ for $k_1=k_2=1$ (bilateral or linear case), with (a) $\veps=0.001$, (b) $\veps=0.003$, (c) $\veps=0.005$,  $L=1$, $a_0=0.01$ and $\omega_1=1.1 \, \omega$, with $\omega$ defined in equation (\ref{frequency}). The effect of the perturbation is to produce an oscillating amplitude at a frequency corresponding to $\omega_1-\omega$. The amplitude of the slower oscillations is proportional to $\veps$. The same behaviour is observed for $k_1=1, k_2=5$ in Figures \ref{fig104}(a-c) and for $k_1=0.001, k_2=1$ in Figures \ref{fig105}(a-c)

\begin{figure}[h!]
  \centering
  \includegraphics[width=4.3cm]{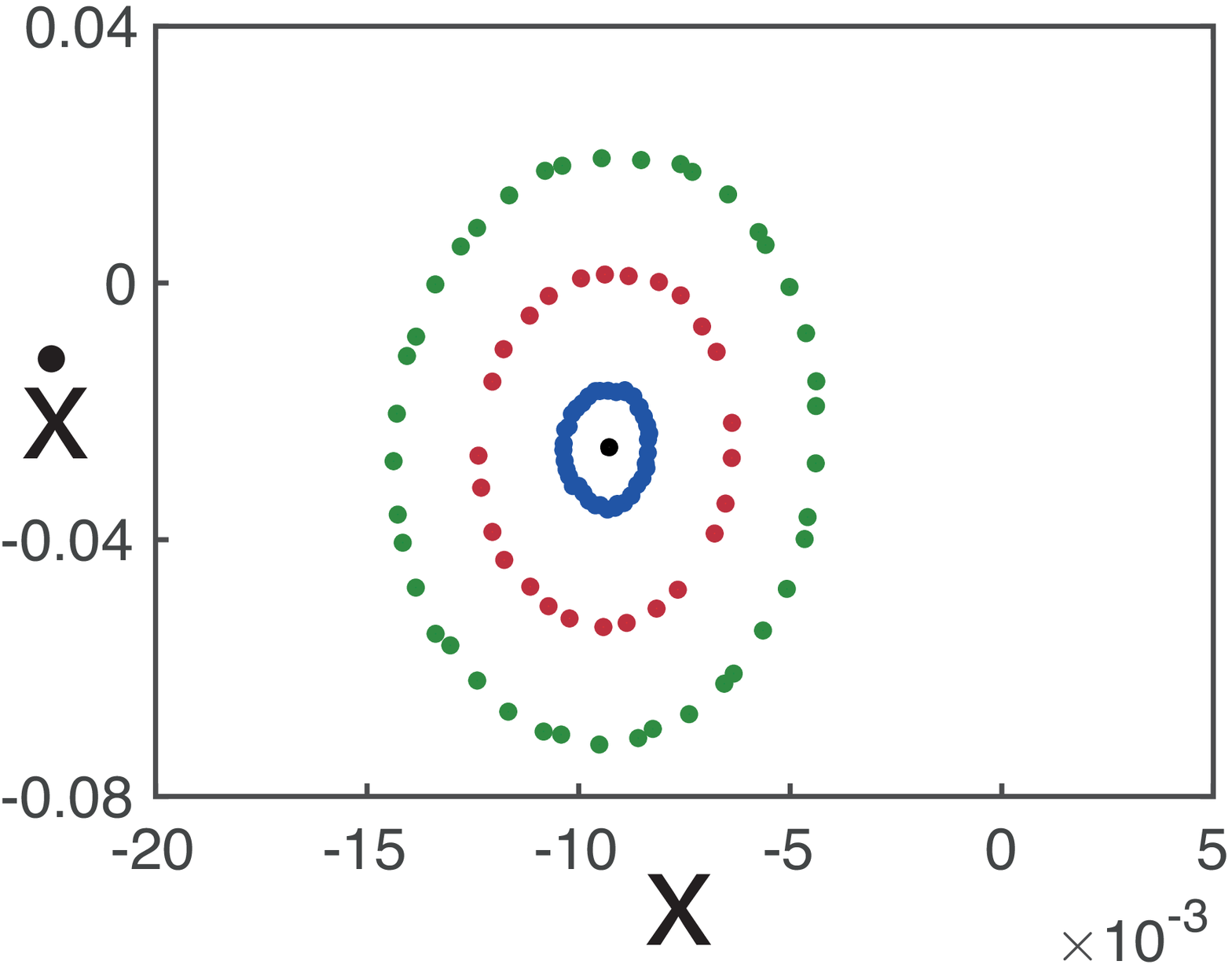}
  \includegraphics[width=4.3cm]{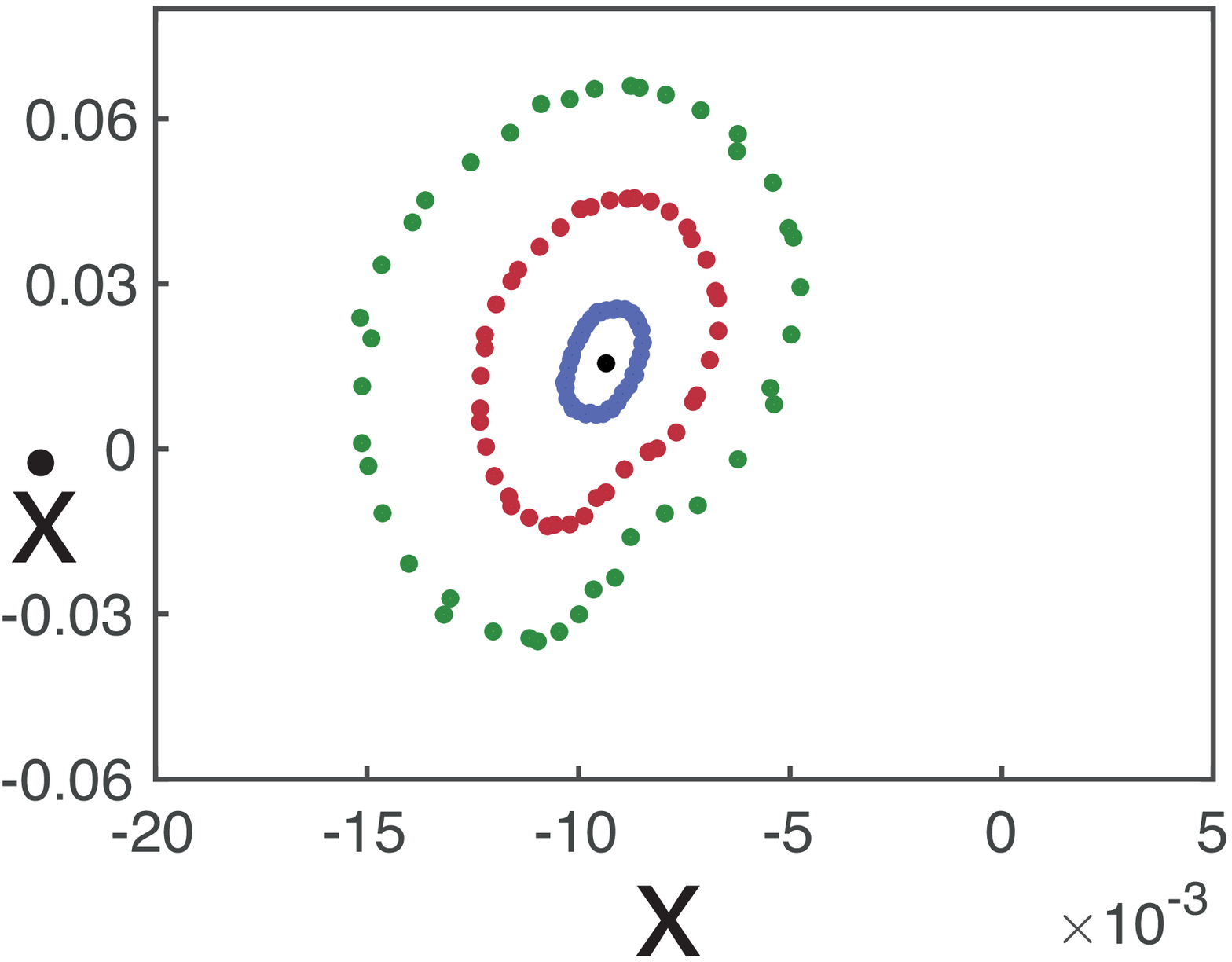}
  \includegraphics[width=4.3cm]{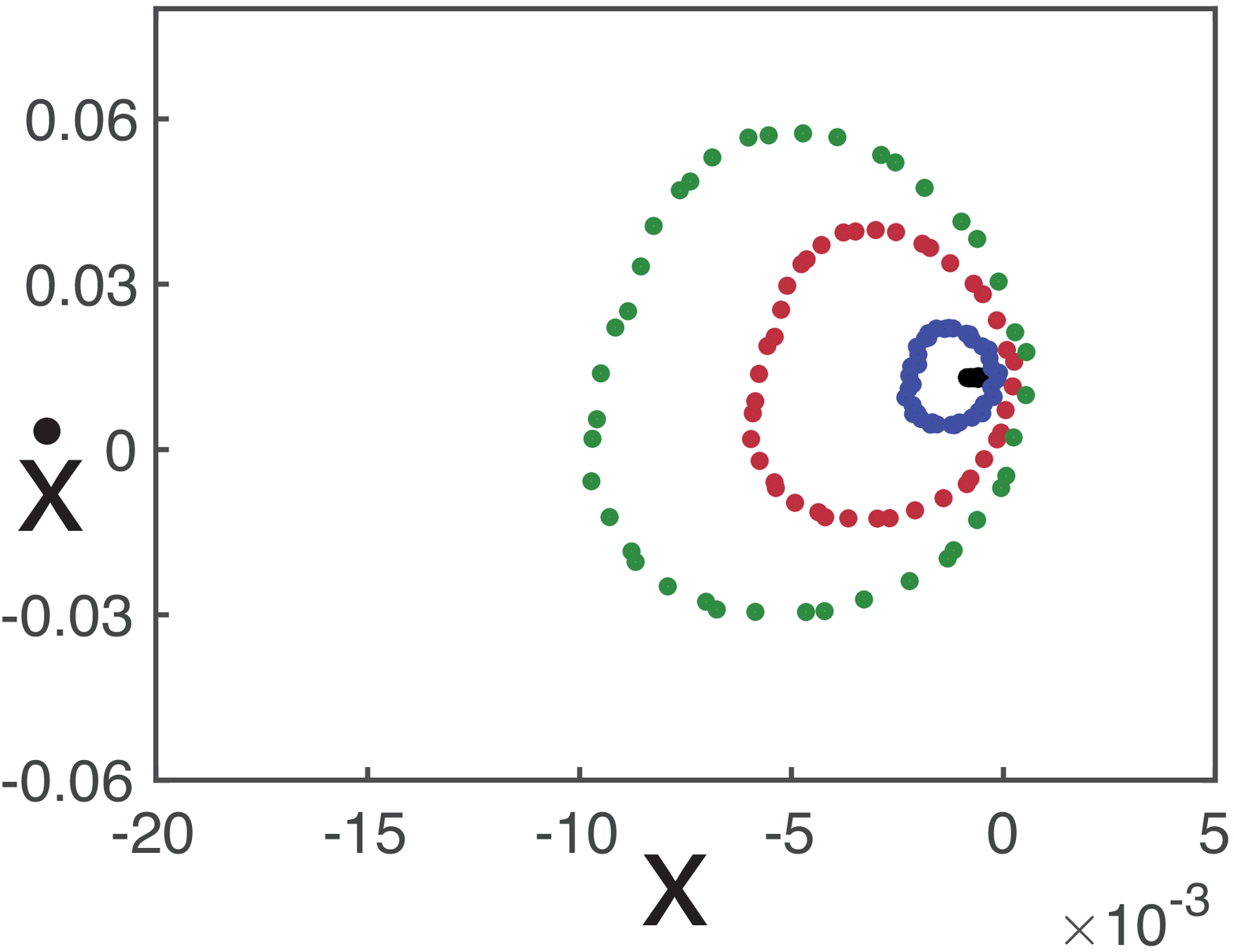}
  \caption{Return map  for $L=1$, $a_0=0.01$, $\omega_1=\sqrt{2} \,\, \omega$, (a) $k_1=k_2=1$ {\corrL (linear case)}, (b) $k_1=1,k_2=5$, (c) $k_1=0.001,k_2=1$  and $\veps=0$ (black dot), $0.001$ (blue dots), $0.003$ (red dots) and $0.005$ (green dots).}
  \label{fig103}
\end{figure}

A convenient representation of the solutions is provided by the map $(f(t),\dot{f}(t)) \to (f(t+\tau),\dot{f}(t+\tau))$ of the phase space $(f(t),\dot{f}(t))$ onto itself, where $f(t)=w(x_0,t)$ with $x_0 > 0$ a suitably chosen point on the simulation domain and $\tau$ the period defined in equation (\ref{frequency}). In rigorous terms, the map is independent of $x_0$. The return maps are shown in Figures \ref{fig103}(a-c) for (a) $k_1=1,k_2=1$, (b) $k_1=1,k_2=5$, (c) $k_1=0.001,k_2=1$. 

The black dot at the center represents the unperturbed solution, the sets of blue, red and green dots around it correspond to $\veps=0.001$, $\veps=0.003$ and $\veps=0.005$, respectively. We see that the phase space points with $\veps \ne 0$ lie on a closed orbit around the unperturbed fixed point, suggesting linear stability, as anticipated in Subsection \ref{ssec:stability}. The different choice of $\omega_1=\sqrt{2} \, \omega$ instead of  $\omega_1=1.1 \, \omega$
adopted in the previous simulations, is dictated by graphical reasons, since it gives a more densely populated phase-space trajectory. {\corrL Note that the closed orbits around the unperturbed fixed point are of elliptical shape in the linear case ($k_1=k_2$) while they appear more and more deformed as the difference between $k_1$ and $k_2$ becomes larger, thus making the nonlinearity more important}.

It is also to be noted that, as $k_1 \to 0$, the orbits surrounding the fixed point tend to become tangent to each other, the periodic point becoming embedded in these curves at their point of contact. This is due to the fact that, as shown before, in this case the solution tends to stay only in the $w<0$ region. \\

When perturbing the initial condition with a non-zero function, we must assure that the perturbed initial condition vanishes on a significant portion of the computational domain near its end, otherwise unwanted waves would propagate inwards from the right boundary.

\begin{figure}[h!]
  \centering
  \includegraphics[width=6 cm]{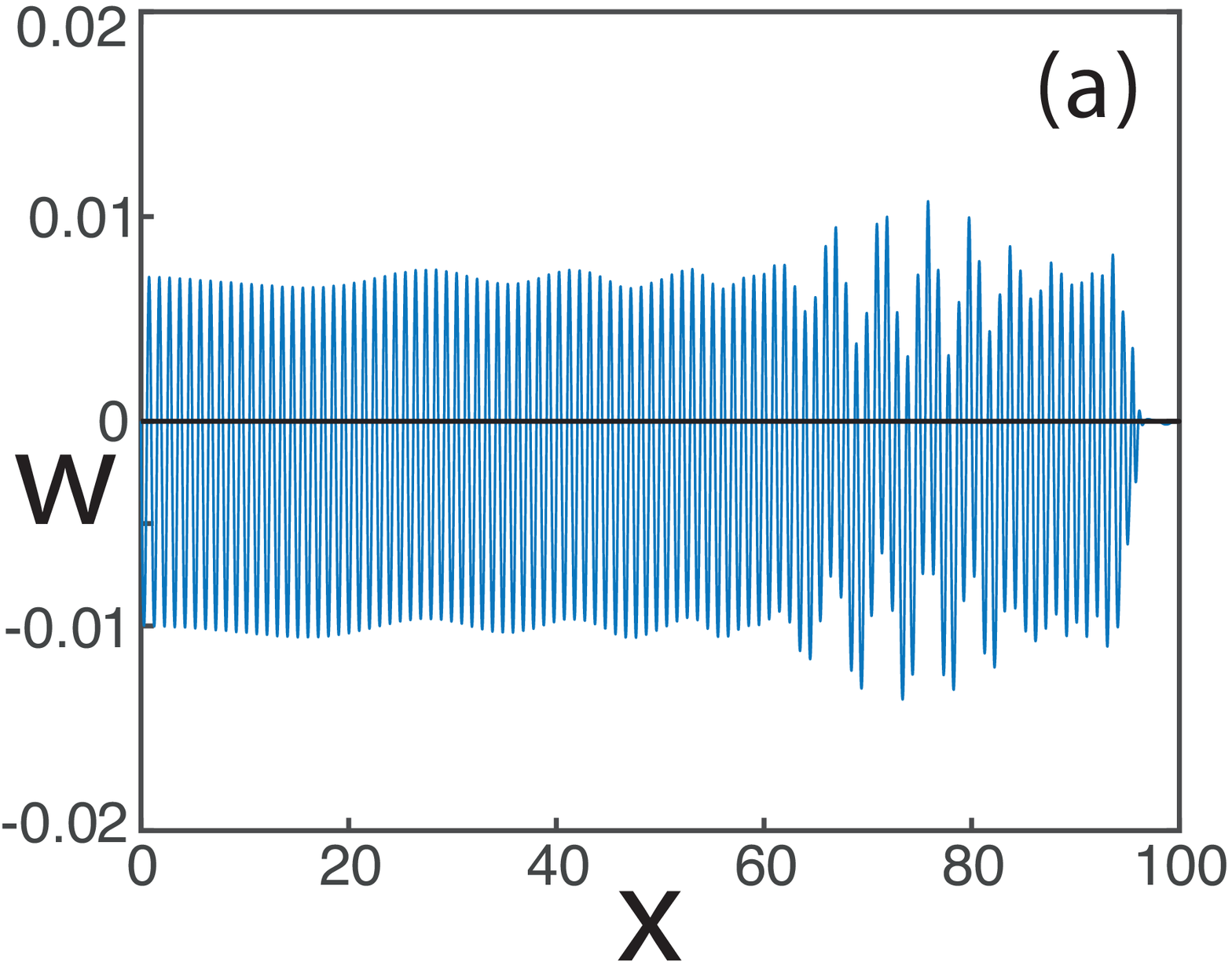} \hspace*{0.5 cm}
  \includegraphics[width=6 cm]{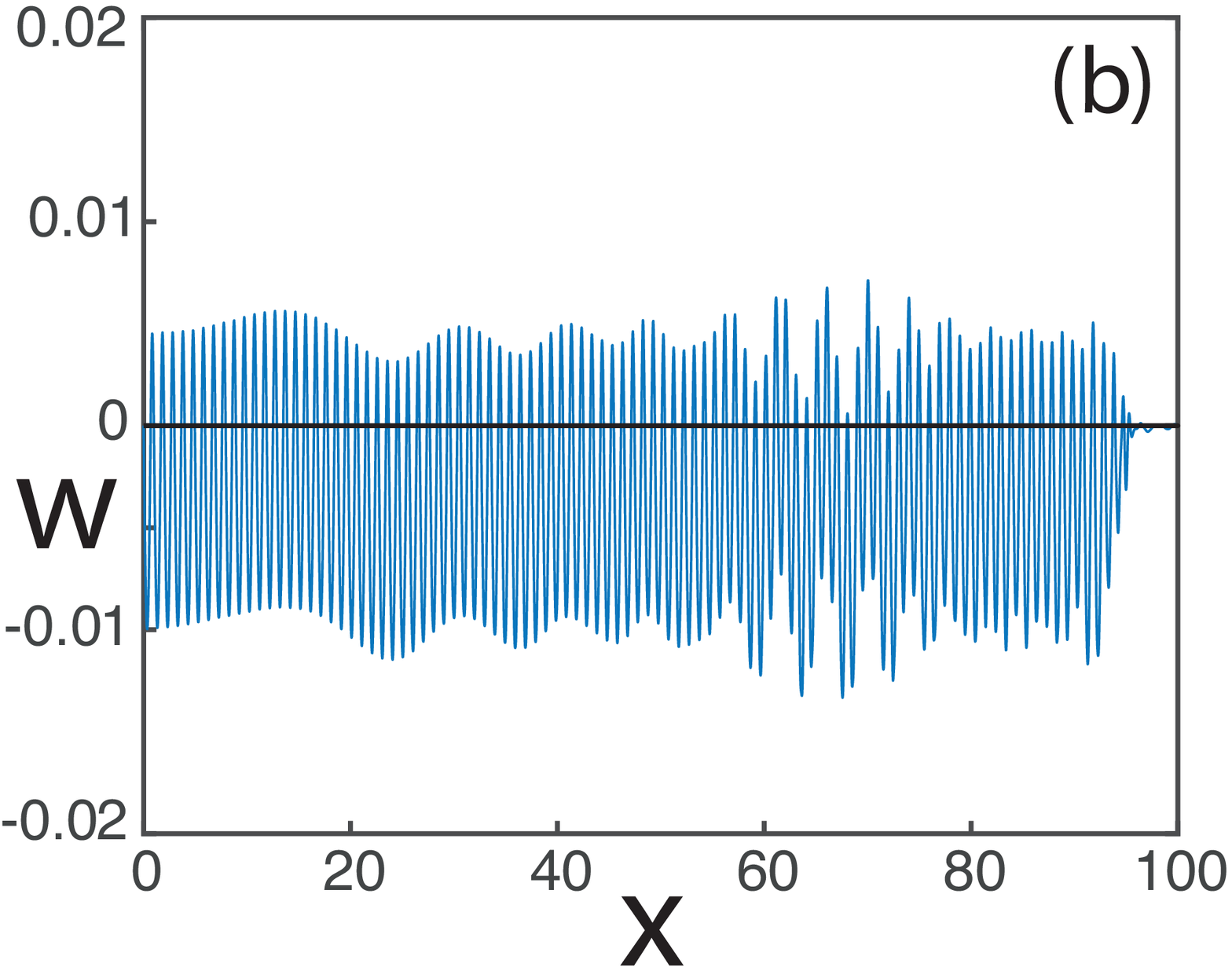}
   \caption{The solution $w(x,t)$ with the initial condition (\ref{PsiNonZero}) for $L=1$, $a_0=0.01$, (a) $k_1=1,k_2=2$, (b) $k_1=1,k_2=5$, and $\veps=0.0003$.}
  \label{fig106}
\end{figure}

\begin{figure}[h!]
  \centering
  \includegraphics[width=6cm]{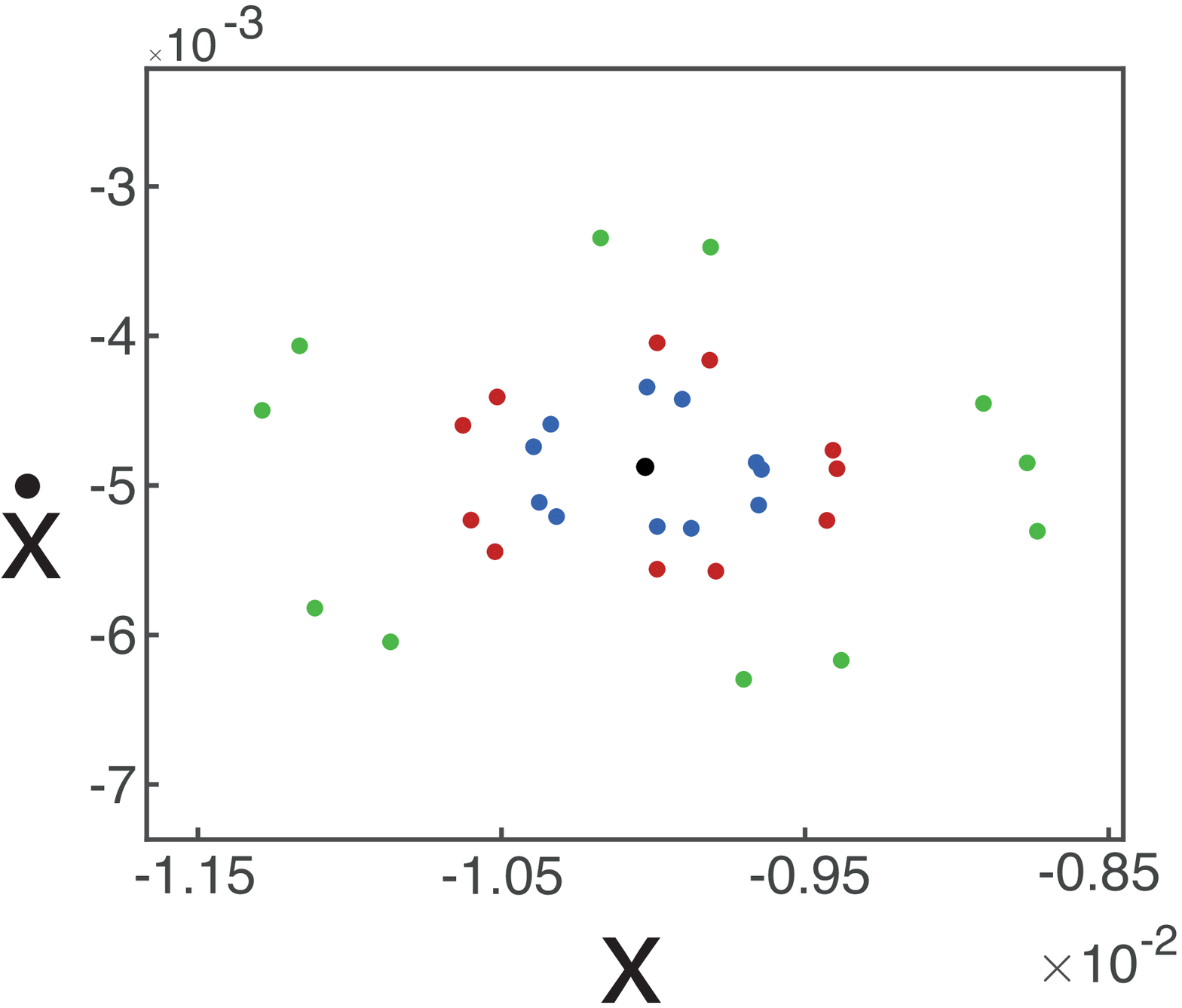} \hspace*{0.5 cm}
  \includegraphics[width=6cm]{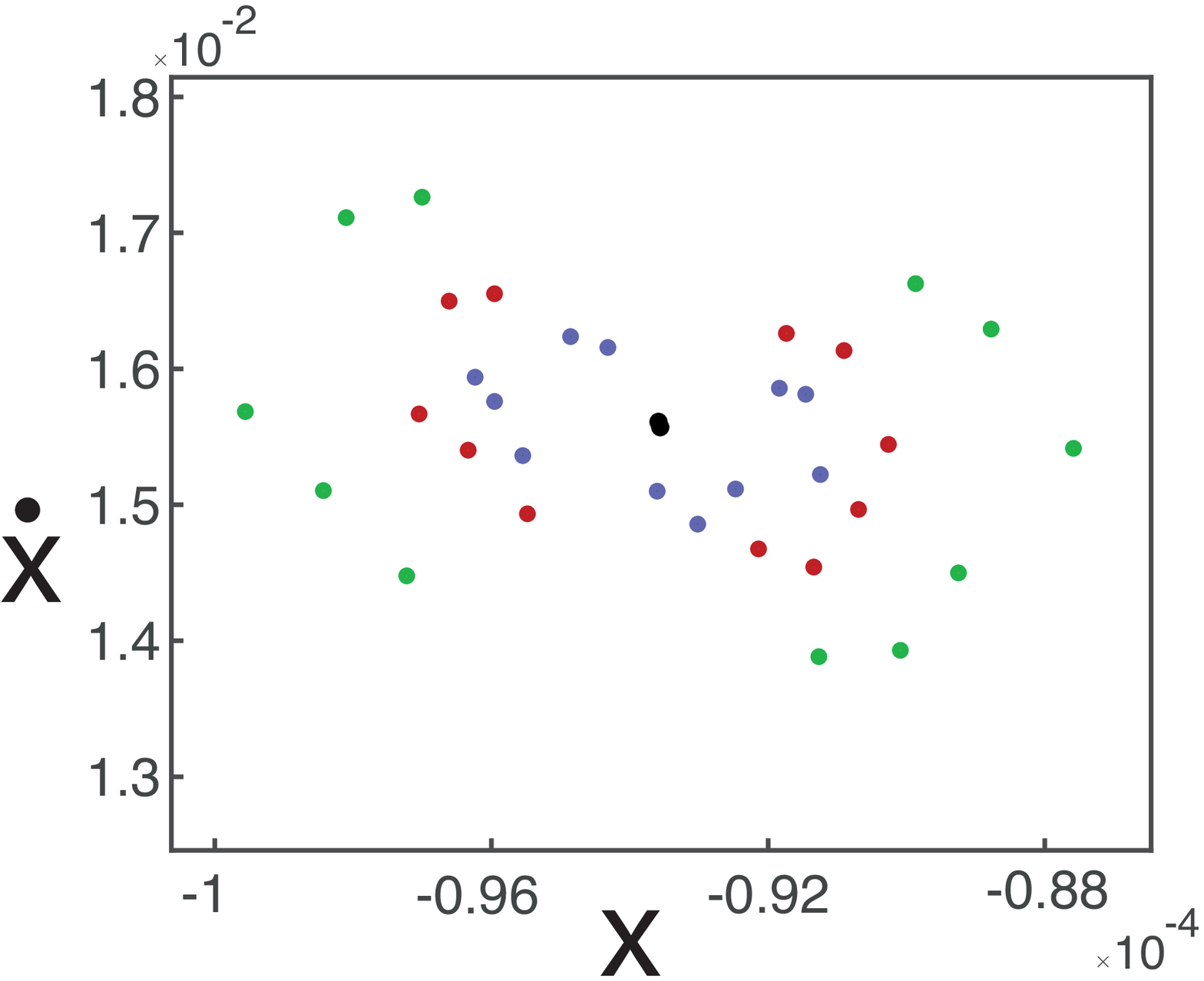}
   \caption{Return map with the initial condition (\ref{PsiNonZero}) for $L=1$, $a_0=0.01$, (a) $k_1=1,k_2=2$, (b) $k_1=1,k_2=5$  and $\veps=0$ (black dot), $0.003$ (blue dots), $0.005$ (red dots) and $0.008$ (green dots).}
  \label{fig107}
\end{figure}

For the example which we illustrate in figures \ref{fig106}(a,b) and \ref{fig107}(a,b) we have 
\beq
w_0(x) = \veps_1 \, \sin(\sqrt{2} \, x) \,\exp(-0.8 \, x). \label{PsiNonZero}
\eeq
and (a) $k_1=1,k_2=2$, (b) $k_1=1,k_2=5$ and let $\veps_1$ vary parametrically as we did before.

The behavior of the solution follows  again the pattern of an initial transient with a subsequent settling to the travelling wave predicted by the model. The transient in this case appears quite longer. The return maps, shown in figures \ref{fig107}(a,b) for $\veps_1=0$, $0.003$,  $0.005$ and $0.008$, indicate again that the solution in linearly stable.

\subsection{With transverse distributed load}

In the presence of a transverse distributed constant load $p$ the governing equations (\ref{eq04aload})-(\ref{eq04bload}) possess a constant particular solution $w_p$ given by
 \begin{align}
& w_p=\frac{p}{k_1}, \quad \mathrm{for} \quad p < 0, \label{wpaload} \\
& w_p=\frac{p}{k_2}, \quad \mathrm{for} \quad p > 0. \label{wpbload}  
\end{align}

It must be remarked that equation (\ref{waveeqload}) possesses many particular solutions, most of which are incompatible with the propagating wave solutions we are dealing with. \\

To search for the numerical solution of equation (\ref{waveeqload}) we subtract the particular solution given by (\ref{wpaload}) or (\ref{wpbload}) and approach numerically only the homogeneous equation with the associated initial - boundary value problem 
\beqa
&& \frac{\partial^2 u}{\partial t^2}  - v^2 \frac{\partial^2 u}{\partial x^2} +\gamma(u+w_0) \, (u+w_p) = 0 \label{kgnlload} \\
&& u(0,t) = \vphi (t)  - w_p \label{kgnlbcload} \\
&& u(x,0) = w_0(x) - w_p; \qquad \frac{\partial u}{\partial t} (x,0) = \psi(x)  \label{kgnlicload}
\eeqa
\begin{figure}[h!]
  \centering
  \includegraphics[width=4.3cm]{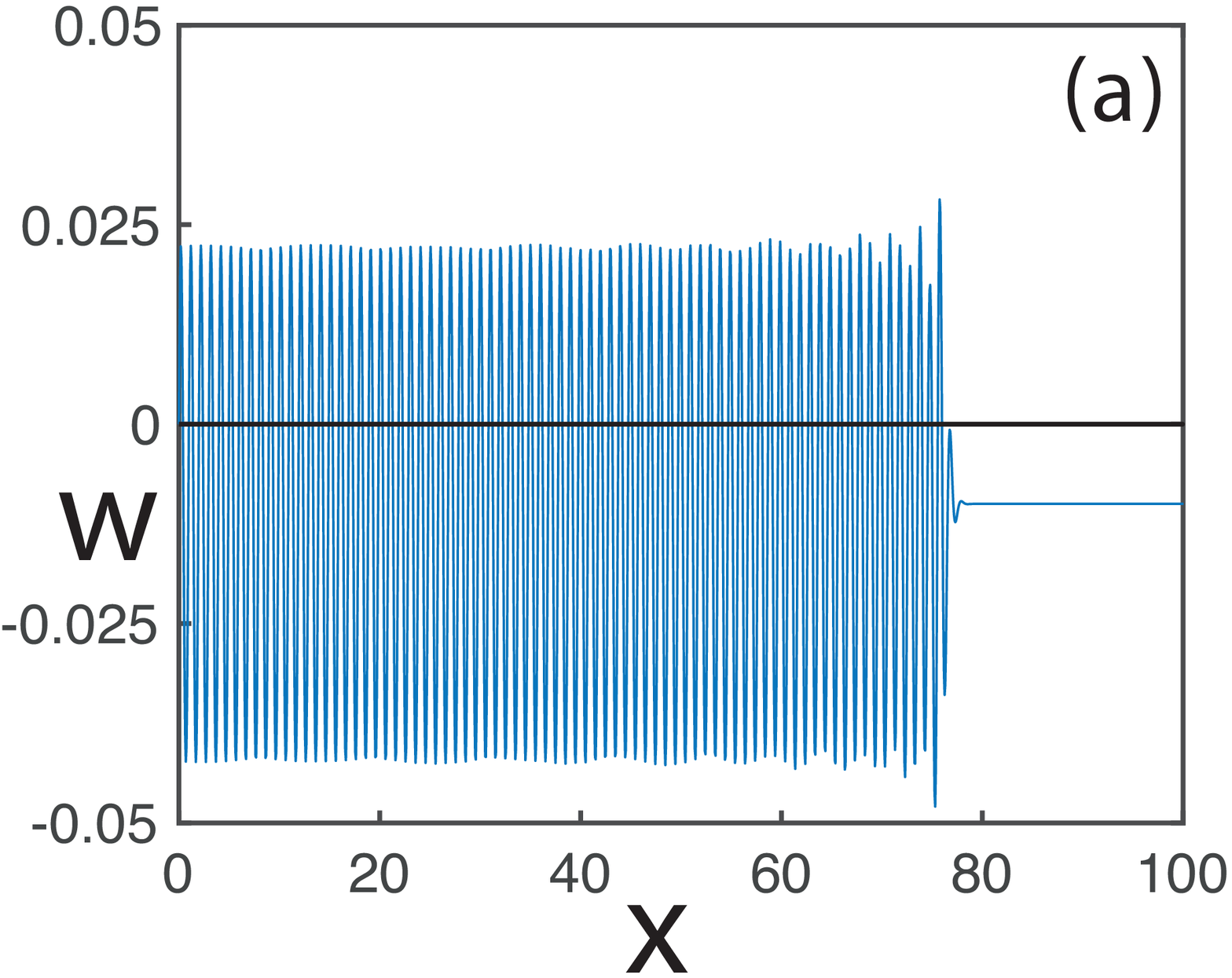}
  \includegraphics[width=4.3cm]{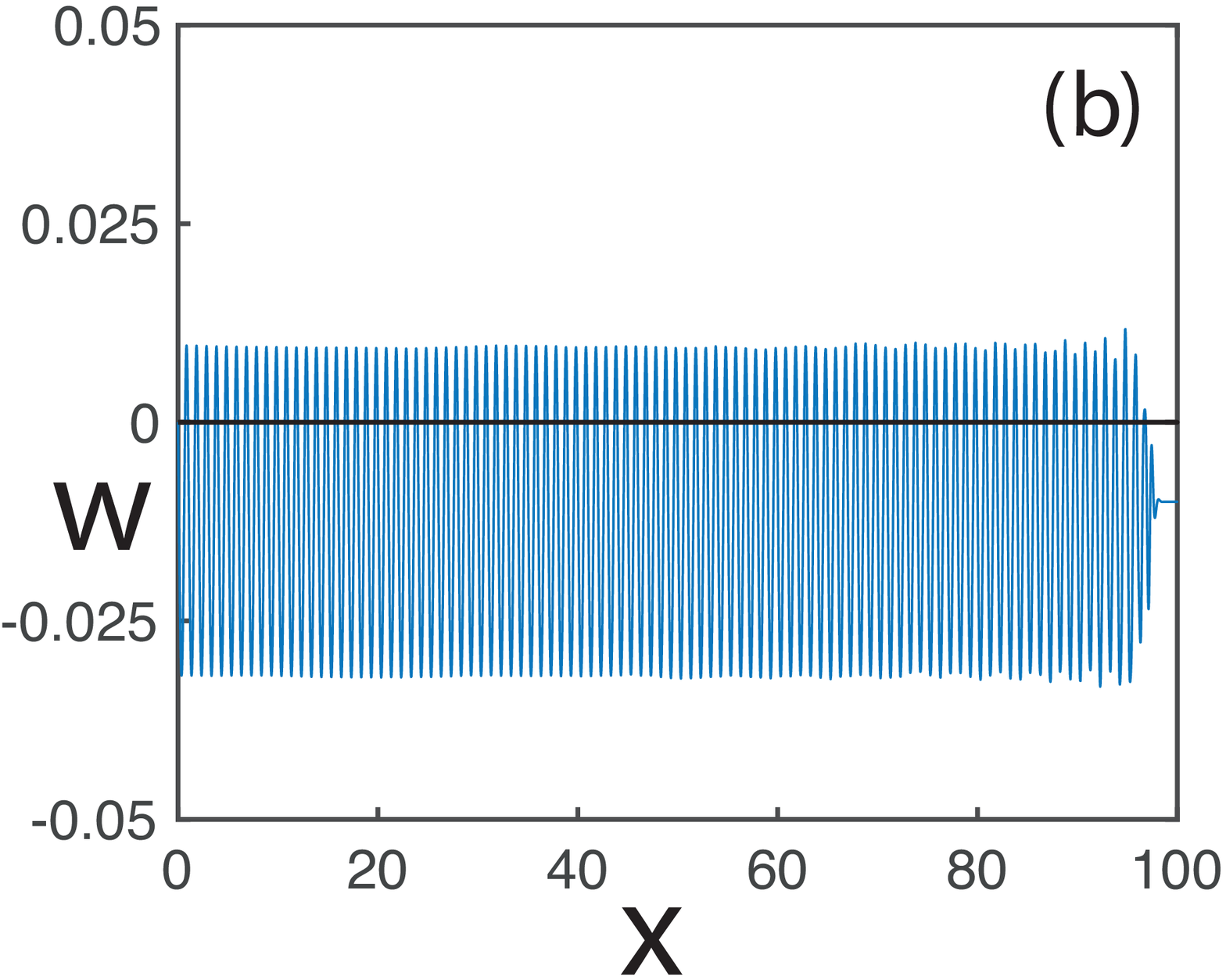}
  \includegraphics[width=4.7cm]{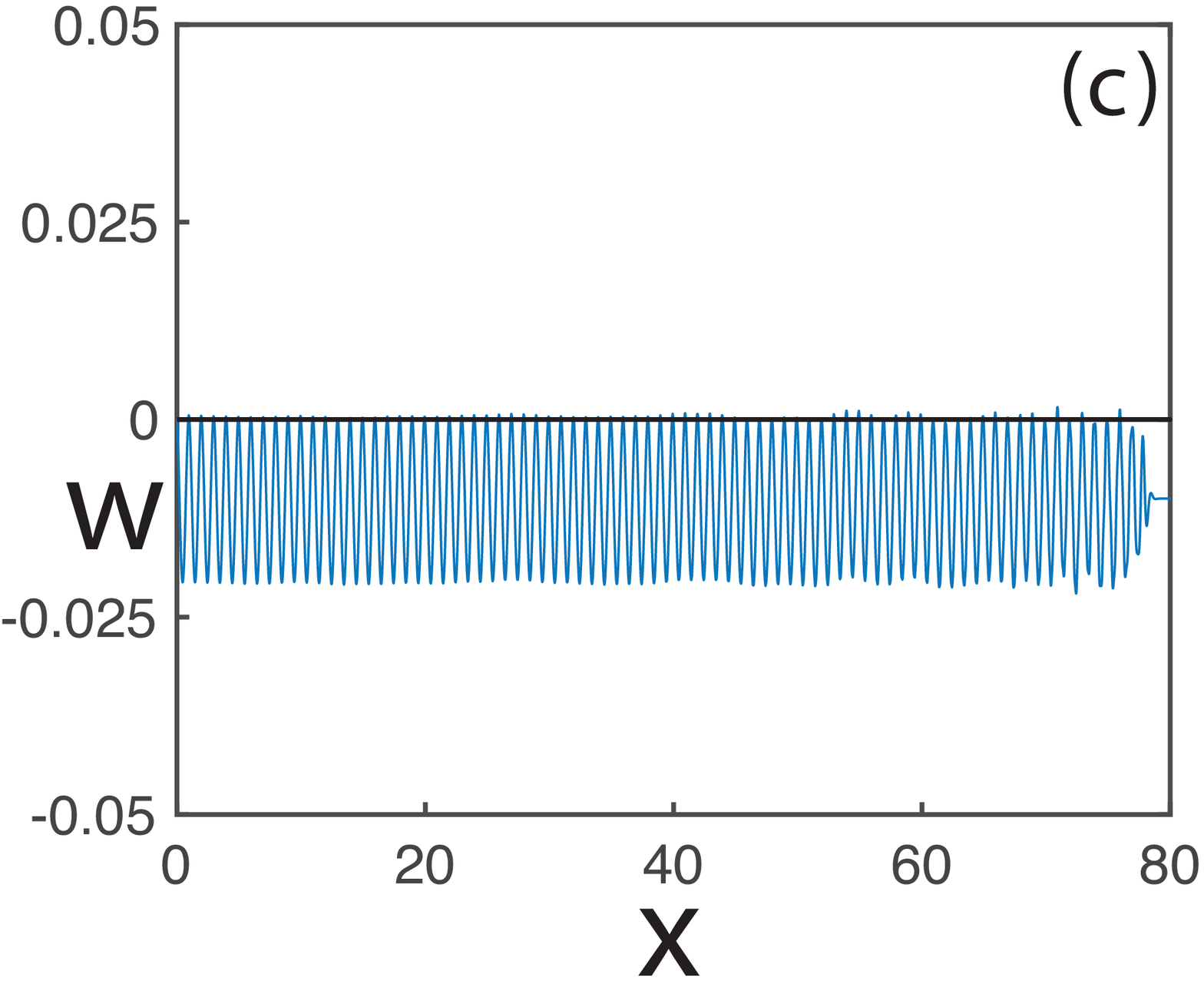}
  \caption{The solution $w(x,t)$ at the final time for (a) $k_1=k_2=1$, (b) $k_1=1,k_2=2$, (c) $k_1=1,k_2=10$ and $L=1$, $p=-0.01$, $\alpha=1.2 \, \alpha_{cr}$.}
    \label{fig200}
\end{figure}
where $u=w-w_p$. The values of $\alpha$ and $a$, which enter the numerical solution via the boundary condition $\vphi(t)$, are given by the solutions of the dispersion relation (\ref{eq23}). As we have remarked in Section \ref{ssec:load}, equation (\ref{eq23}) possesses a multiplicity of solutions for any given $\alpha$. It turns out that only the lowest solution corresponds to a simple wave, while the higher ones correspond to multiple waves. 
\begin{figure}[h!]
  \centering
  \includegraphics[width=12cm]{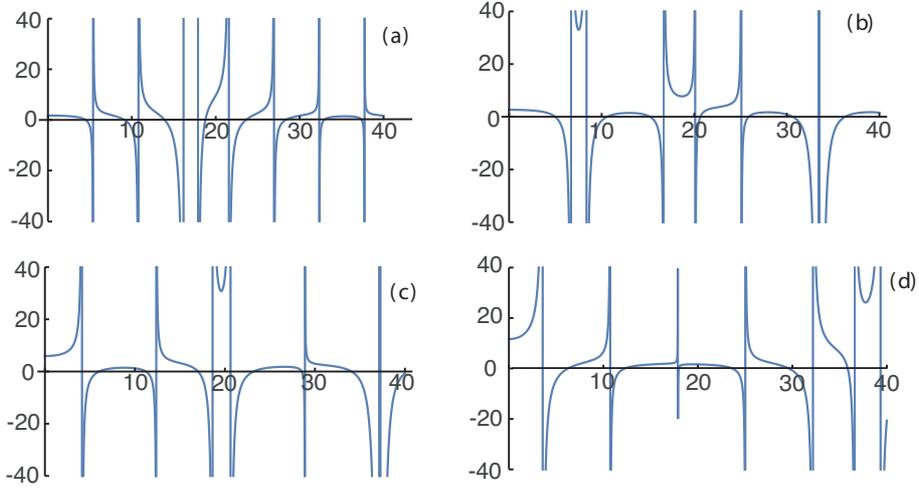}
  \caption{The dispersion relation (\ref{eq23}) as function of $a$  for $k_1=1,k_2=2$ and (a) $\alpha=0.3 \, \alpha_{cr}$, (b) $\alpha=0.8 \, \alpha_{cr}$, (c) $\alpha=1.3 \, \alpha_{cr}$ (d) $\alpha=1.8 \, \alpha_{cr}$ .}
  \label{figDispRel}
\end{figure}
The dispersion relation (\ref{eq23}) is shown in Fig. \ref{figDispRel}(a-d) for $k_1=1,k_2=2$ and four different values of $\alpha$; here $\alpha_{cr}=0.585786$. The solutions, which give the value of $a$ for a given $\alpha$, form a countably infinite set and only the lowest one correspond to a simple wave. The first singularity corresponds to $a=\pi/\alpha_{cr}$; note how, by varying $\alpha$, the position of the roots with respect to the singularities changes. In this work we are interested only in simple waves, that is in the lowest roots of the dispersion relation. The presence of multiple roots, however, entails that the waves corresponding to the higher roots coexist with the fundamental one. As a consequence, the propagating traveling waves described by our model cannot be reproduced exactly by the numerical solution, because the numerical errors introduced by the discretization of the equations excite these higher waves. Because of this problem, the constant amplitude of the traveling wave presents a slight modulation. \\
\begin{figure}
  \centering
 \includegraphics[width=6cm]{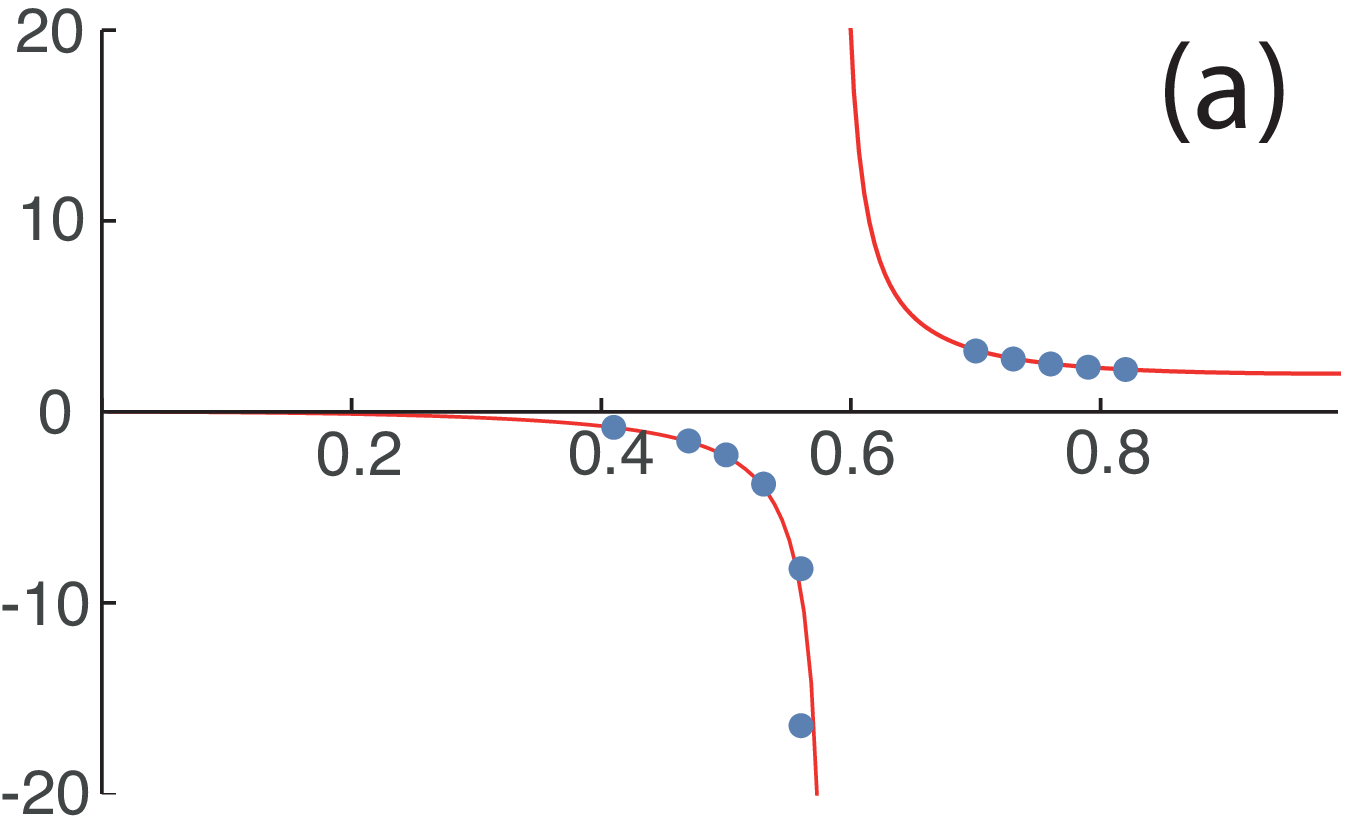} \hspace*{1 cm}
\includegraphics[width=6cm]{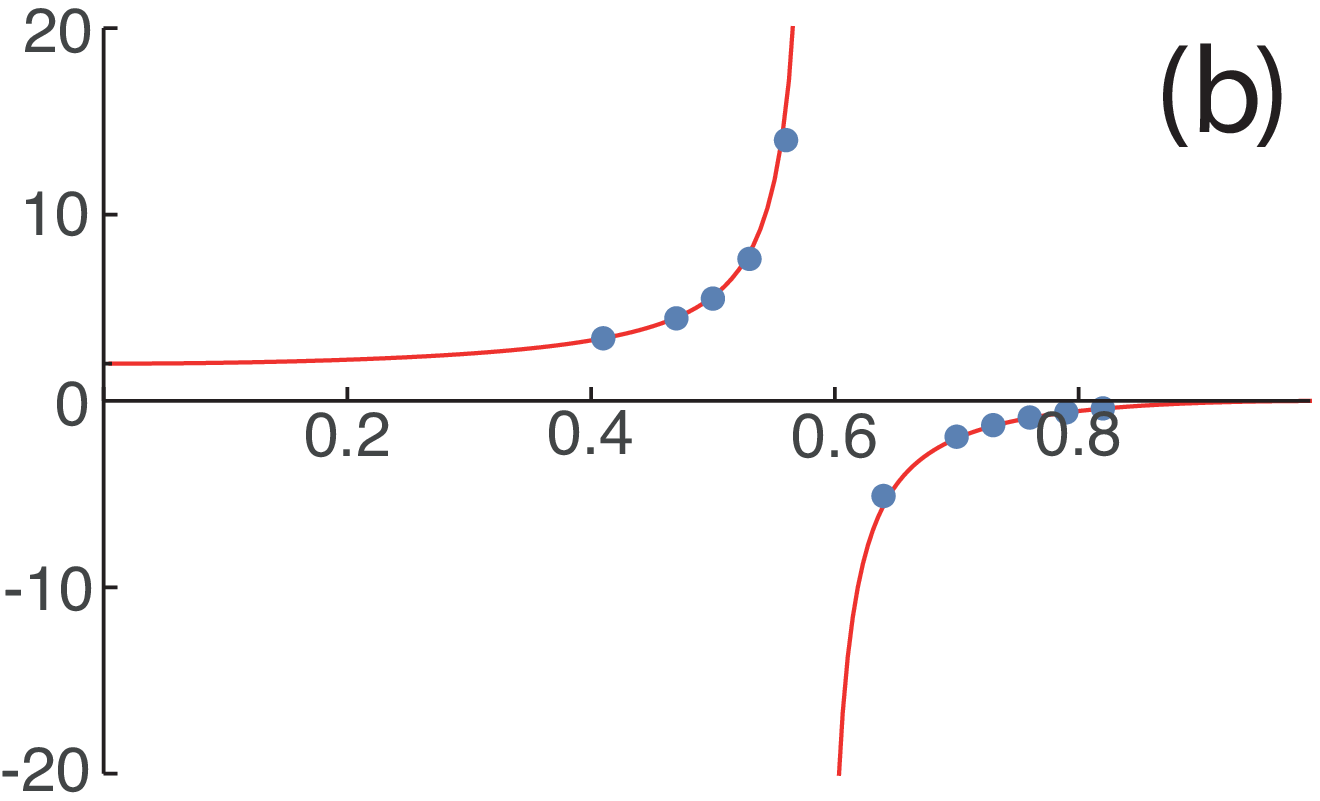}
  \caption{The maximum and miminum of $W(s)$ for $k_1=1,k_2=2$,  varying 
  $0 \le \alpha \le 1$. Here 
  $\alpha_{cr}=0.585786$.}
  \label{fig05}
\end{figure}

The propagating wave solution $w(x,t)$ is portrayed in Figures \ref{fig200}(a-c) for $p=-0.01$, $L=1$, $\alpha=1.2 \, \alpha_{cr}$ and (a) $k_1=k_1=1$, (b) $k_1=1,k_2=2$,  (c) $k_1=1,k_2=10$. In all three cases, we observe a similar behaviour as in the case without the transverse load: after an initial transient, the propagating wave settles to the right-propagating wave with constant amplitude, which is the restriction to $x \ge 0$ of the wave found analytically in Section \ref{ssec:load}. \\

In Fig. \ref{fig05}(a,b) we highlight the role of $\alpha_{cr}$ by showing the maximum excursions of $W_1$ and $W_2$, normalized to $k_1/p$ and $k_2/p$ respectively, as functions of $\alpha$ for $k_1=1$ and $k_2=2$, which results in $\alpha_{cr} = 0.585786$; the singularity in the amplitude at $\alpha=\alpha_{cr}$ and the requirement on the sign of the transversal load $p$ are evident (see the discussion following equation (\ref{eq23})). In these figures, the red lines represent the theoretical result, obtained from equations (\ref{eq21a})  and (\ref{eq21b}) with the coefficients given by (\ref{eq22}) with $a$ and $\alpha$ related by the dispersion relation (\ref{eq23}). The blue dots are the amplitudes calculated from the numerical solution; the agreement is excellent when $\alpha$ is far from $\alpha_{cr}$, while for $\alpha$ close to $\alpha_{cr}$  the discretization errors become somewhat larger. \\

\begin{figure}[h!]
  \centering
  \includegraphics[width=4.3cm]{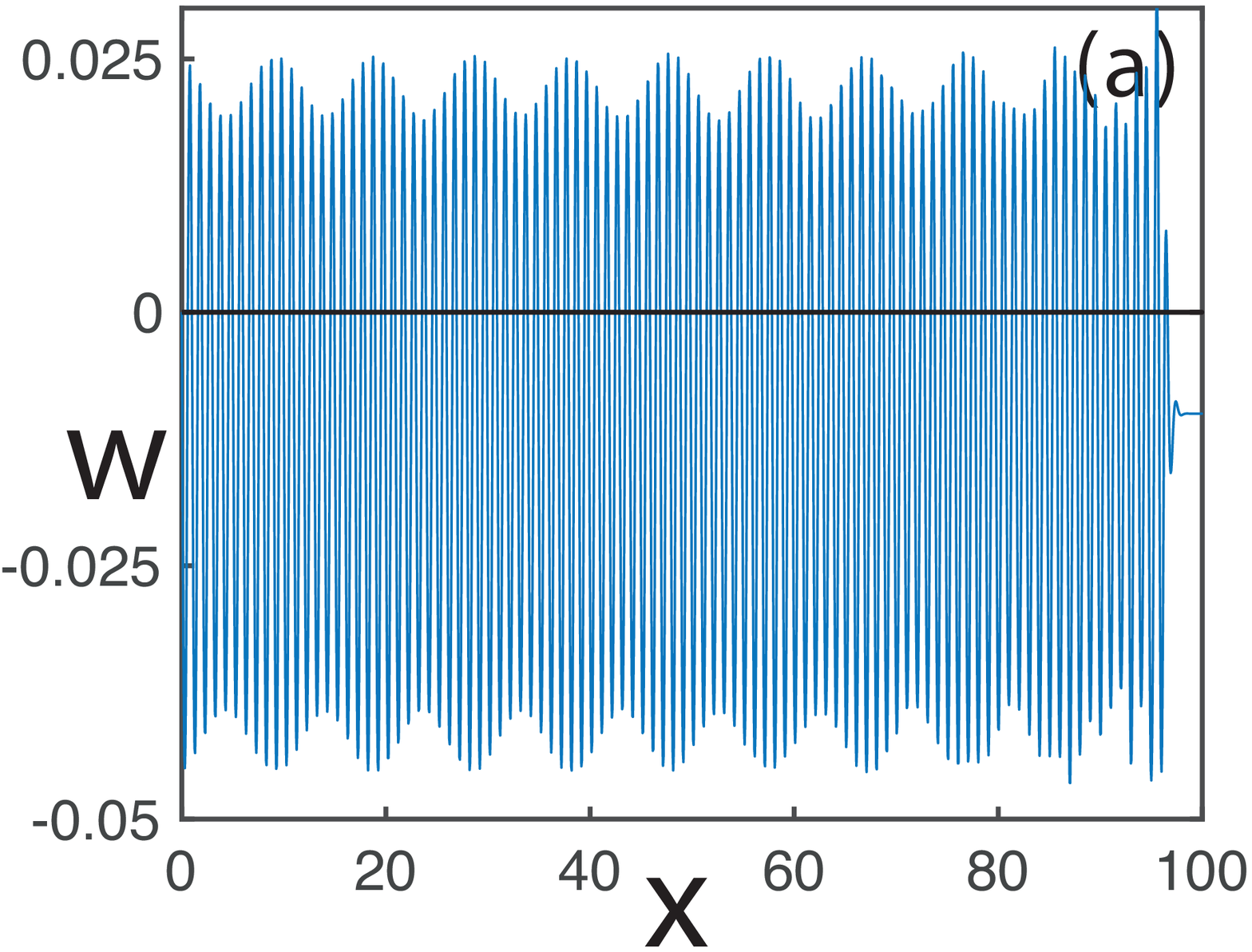} \hspace*{0.2cm}
  \includegraphics[width=4.3cm]{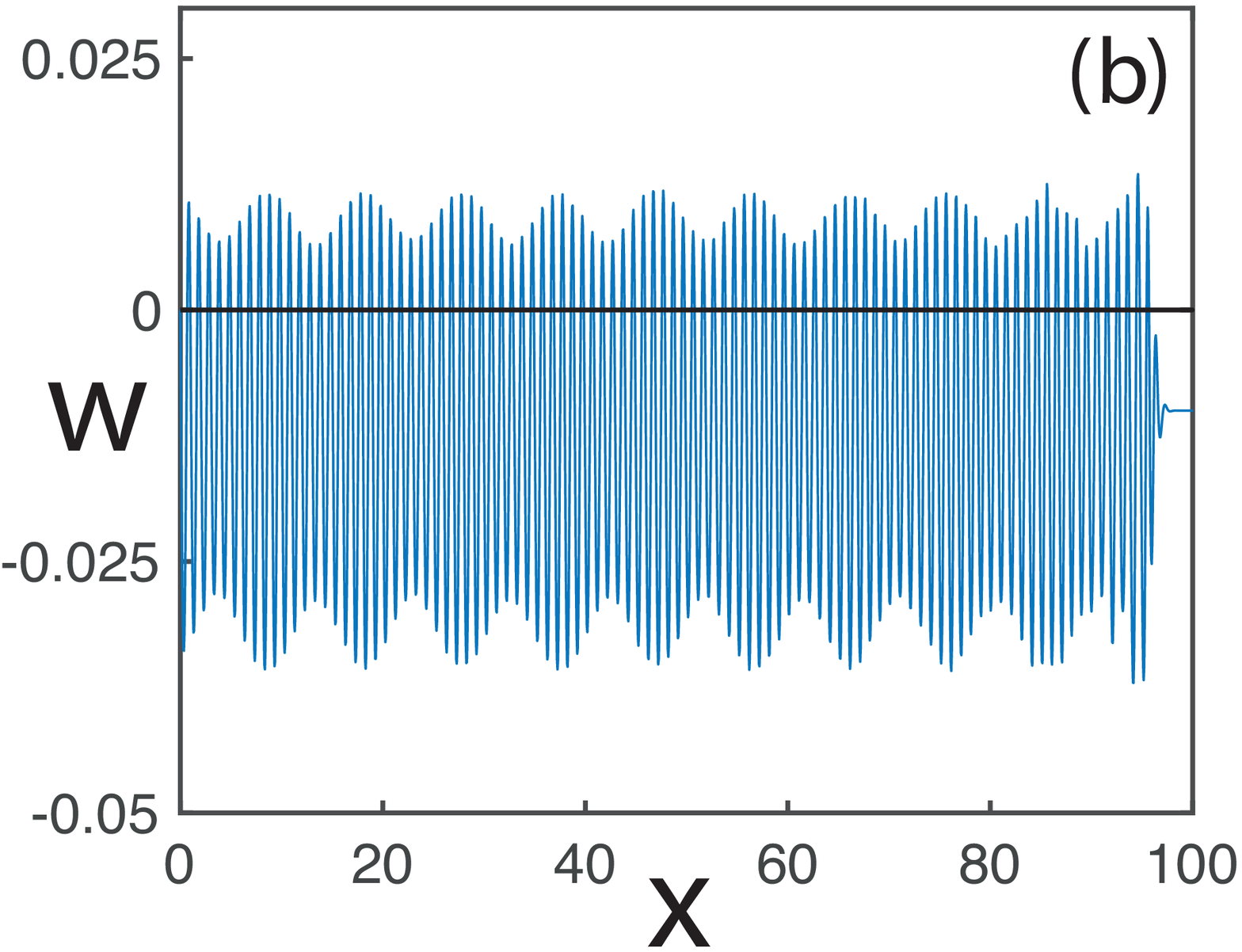} \hspace*{0.2cm}
  \includegraphics[width=4.1cm]{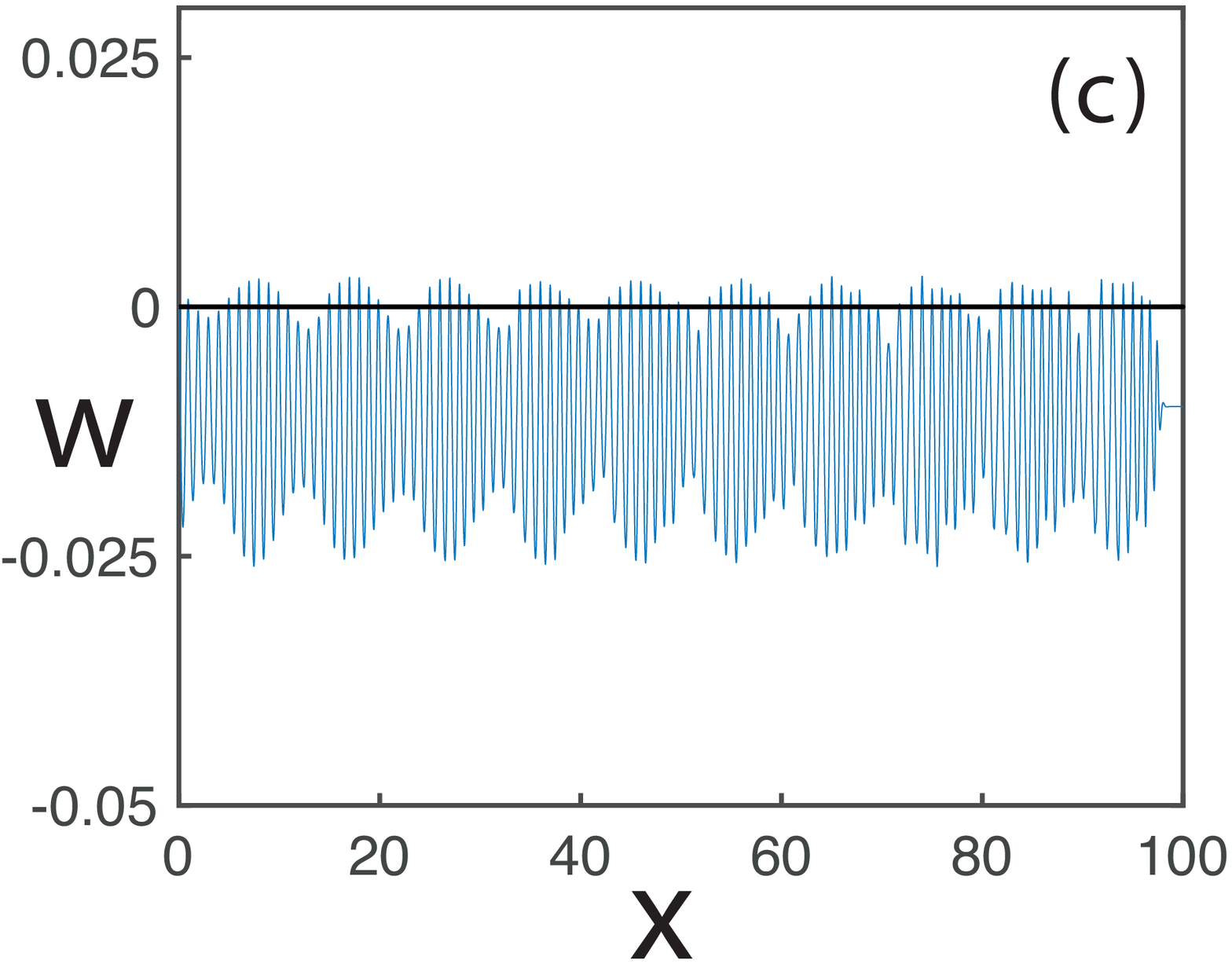}
  \caption{The solution $w(x,t)$ at the final time for (a) $k_1=k_2=1$ {\corrL (linear case)}, (b) $k_1=1,k_2=2$, (c) $k_1=1,k_2=10$ and $L=1$, $p=-0.01$, $\alpha=1.2 \, \alpha_{cr}$, 
  $\veps=0.003$ and $\omega_1=1.1 \, \omega$.}
  \label{fig201}
\end{figure}

Along the same lines followed in the homogeneous case, we study numerically the stability of the traveling wave solutions by introducing a small perturbation of the form (\ref{perturb}) in the boundary condition. The solution is shown in 
figure \ref{fig201}(a-c) for (a) $k_1=k_1=1$, (b) $k_1=1,k_2=2$ and (c) $k_1=1,k_2=10$, $L=1$, $p=-0.01$, $\alpha=1.2 \, \alpha_{cr}$. $\veps=0.003$ and $\omega_1=1.1 \, \omega$. We observe the same behaviour seen in the case without load, namely an oscillating amplitude at a frequency corresponding to $\omega_1-\omega$. The amplitude of the slower oscillations is proportional to $\veps$. We obtain the same qualitative scenario for a wide range of parameters, $k_1$ and $k_2$, $p$, $\alpha$ and $\omega_1$.  

\begin{figure}[h!]
  \centering
  \includegraphics[width=4.3cm]{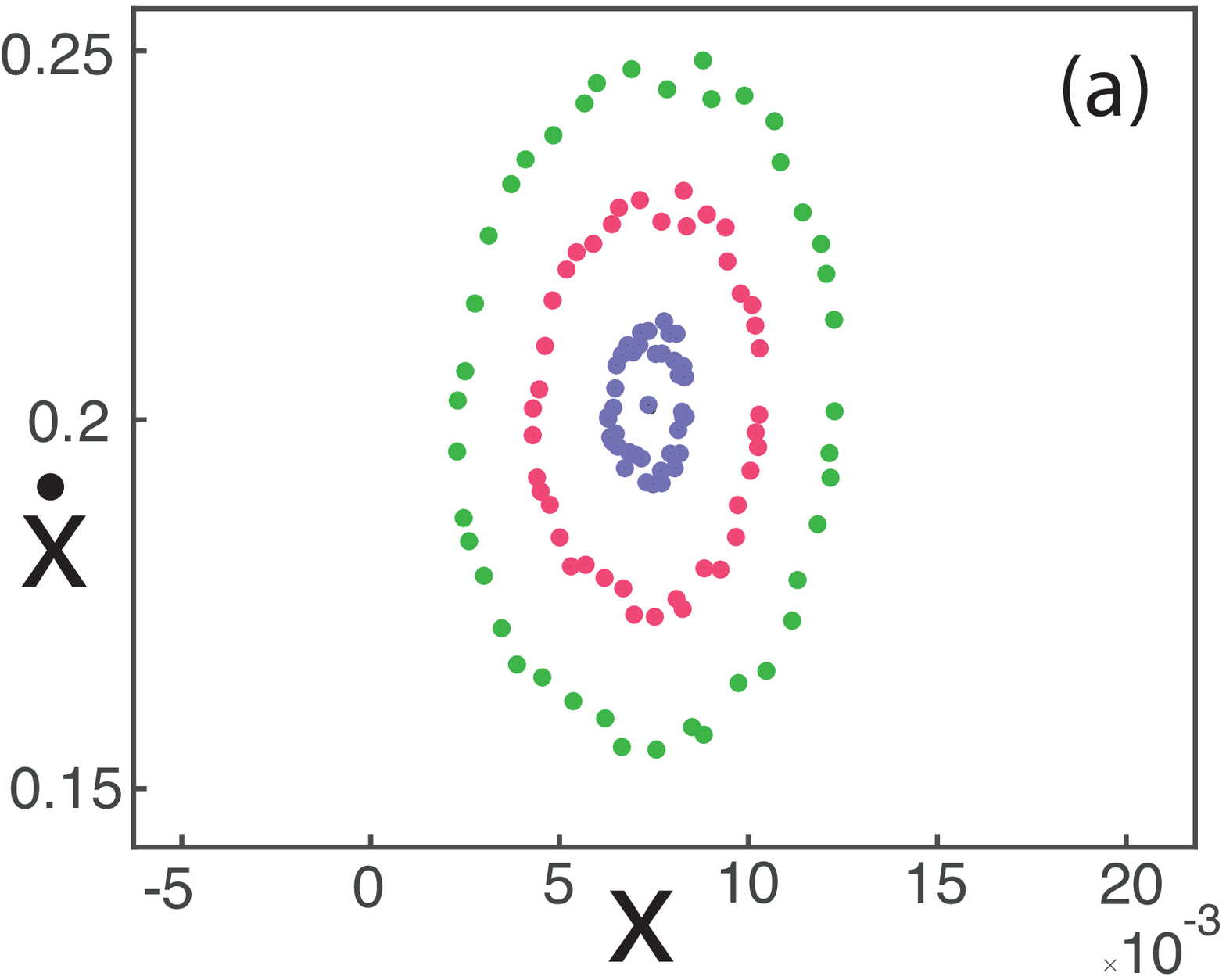}\hspace*{0.2 cm}
  \includegraphics[width=4.3cm]{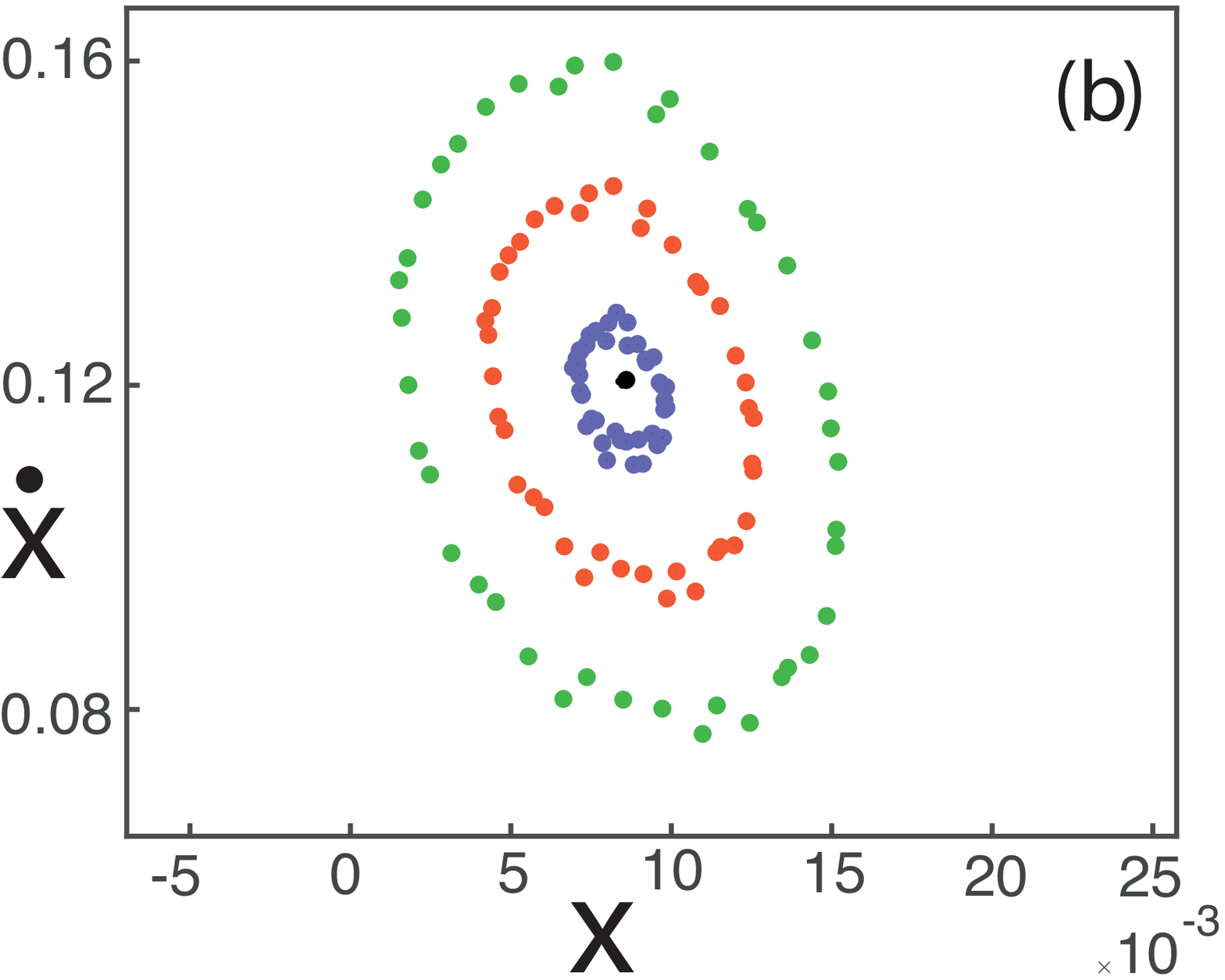}\hspace*{0.2 cm}
  \includegraphics[width=4.3cm]{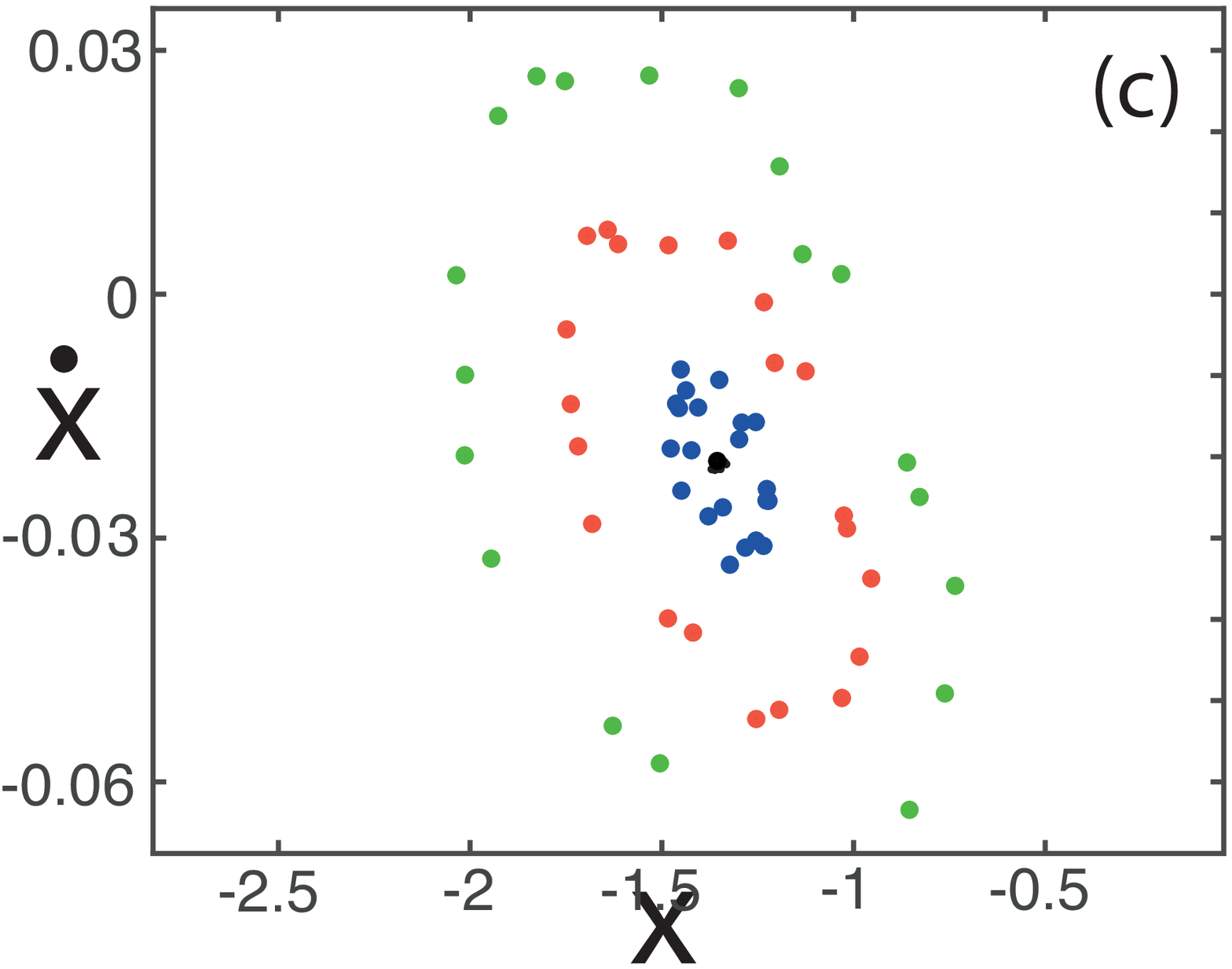}
   \caption{Return map  (a) $k_1=k_2=1$ {\corrL (linear case)}, (b) $k_1=1,k_2=2$, (c) $k_1=1,k_2=5$, and $L=1$, $p=-0.01$, $\alpha=1.2 \, \alpha_{cr}$, 
 and $\omega_1=\sqrt{2} \, \omega$  and $\veps=0$ (black dot), $0.001$ (blue dots), $0.003$ (red dots) and $0.005$ (green dots).}
  \label{fig202}
\end{figure}
Also in this case it is useful to show the return maps $(f(t),\dot{f}(t)) \to (f(t+\tau),\dot{f}(t+\tau))$. In figure \ref{fig202} we have (a) $k_1=k_2=1$, (b) $k_1=1,k_2=2$, (c) $k_1=1,k_2=5$ with $L=1$, $p=-0.01$, $\alpha=1.2 \, \alpha_{cr}$, 
 and $\omega_1=\sqrt{2} \, \omega$  and $\veps=0$ (black dot), $0.001$ (blue dots), $0.003$ (red dots) and $0.005$ (green dots).  Again, the phase space points with $\veps \ne 0$ lie on  closed orbits around the unperturbed fixed point, indicating that the periodic traveling waves are stable also when a transversal distributed load is applied. \\

\section{Conclusions}
\label{conclusions}
We have investigated theoretically and numerically the occurrence of periodic traveling waves supported by the wave equation with the addition of a nonlinear piecewise constant stiffness. In particular, we have studied periodic simple waves, which possess only one node internal to the periodicity interval. In the absence of an external distributed load, the propagation speed and the spatial extension of the two subintervals (a compression interval of size $\alpha$, where the solution is negative and a tension interval of size $1-\alpha$, where the solution is positive) are determined directly as functions of the values of the stiffnesses, while the amplitude of the waves in undetermined. When a distributed load is added, the size of the compression interval $\alpha$ becomes a free parameter, while the propagation speed is determined by the solution of a dispersion relation. The amplitude is then a function of $\alpha$ and it exhibits a singularity when $\alpha$ equals the value of the case without load. \\

The properties of the solutions, without and with the external distributed load, have been verified by a direct numerical solution of the wave equation, obtained by a simple finite difference approach. The stability of these waves has also been studied by numerical means. We have introduced both time-histories and return maps; all our results indicate that the periodic travelling wave solutions are linearly stable, at least against the two kinds of perturbations considered here. \\

{\corrL The nonlinearity introduced in the model by assumptions (\ref{stiffness1})-(\ref{stiffness2}) when $k_1 \ne k_2$  produces the asymmetry of the wave profile with respect to the $w=0$ baseline, as seen in the numerical simulations; it becomes more important in the limiting cases of unilateral and unilaterally rigid substrate. The nonlinearity causes also the deformation of the return-map orbits from the elliptical shape seen in the linear case ($k_1=k_2$).}


\end{document}